\documentclass[twocolumn]{aastex631}
\usepackage[T5,T1]{fontenc}
\pdfoutput=1 %for arXiv submission
\usepackage{amsmath,amstext}
\usepackage{apjfonts} 
\usepackage[figure,figure*]{hypcap}
\usepackage{cleveref}
\usepackage{color}

\shorttitle{IceCube search for CCSNe}
\shortauthors{R. Abbasi et al}

\begin{document}

\title{Search for Galactic core-collapse supernovae in a decade of data taken with the IceCube Neutrino Observatory}

\affiliation{III. Physikalisches Institut, RWTH Aachen University, D-52056 Aachen, Germany}
\affiliation{Department of Physics, University of Adelaide, Adelaide, 5005, Australia}
\affiliation{Dept. of Physics and Astronomy, University of Alaska Anchorage, 3211 Providence Dr., Anchorage, AK 99508, USA}
\affiliation{Dept. of Physics, University of Texas at Arlington, 502 Yates St., Science Hall Rm 108, Box 19059, Arlington, TX 76019, USA}
\affiliation{CTSPS, Clark-Atlanta University, Atlanta, GA 30314, USA}
\affiliation{School of Physics and Center for Relativistic Astrophysics, Georgia Institute of Technology, Atlanta, GA 30332, USA}
\affiliation{Dept. of Physics, Southern University, Baton Rouge, LA 70813, USA}
\affiliation{Dept. of Physics, University of California, Berkeley, CA 94720, USA}
\affiliation{Lawrence Berkeley National Laboratory, Berkeley, CA 94720, USA}
\affiliation{Institut f{\"u}r Physik, Humboldt-Universit{\"a}t zu Berlin, D-12489 Berlin, Germany}
\affiliation{Fakult{\"a}t f{\"u}r Physik {\&} Astronomie, Ruhr-Universit{\"a}t Bochum, D-44780 Bochum, Germany}
\affiliation{Universit{\'e} Libre de Bruxelles, Science Faculty CP230, B-1050 Brussels, Belgium}
\affiliation{Vrije Universiteit Brussel (VUB), Dienst ELEM, B-1050 Brussels, Belgium}
\affiliation{Department of Physics and Laboratory for Particle Physics and Cosmology, Harvard University, Cambridge, MA 02138, USA}
\affiliation{Dept. of Physics, Massachusetts Institute of Technology, Cambridge, MA 02139, USA}
\affiliation{Dept. of Physics and The International Center for Hadron Astrophysics, Chiba University, Chiba 263-8522, Japan}
\affiliation{Department of Physics, Loyola University Chicago, Chicago, IL 60660, USA}
\affiliation{Dept. of Physics and Astronomy, University of Canterbury, Private Bag 4800, Christchurch, New Zealand}
\affiliation{Dept. of Physics, University of Maryland, College Park, MD 20742, USA}
\affiliation{Dept. of Astronomy, Ohio State University, Columbus, OH 43210, USA}
\affiliation{Dept. of Physics and Center for Cosmology and Astro-Particle Physics, Ohio State University, Columbus, OH 43210, USA}
\affiliation{Niels Bohr Institute, University of Copenhagen, DK-2100 Copenhagen, Denmark}
\affiliation{Dept. of Physics, TU Dortmund University, D-44221 Dortmund, Germany}
\affiliation{Dept. of Physics and Astronomy, Michigan State University, East Lansing, MI 48824, USA}
\affiliation{Dept. of Physics, University of Alberta, Edmonton, Alberta, Canada T6G 2E1}
\affiliation{Erlangen Centre for Astroparticle Physics, Friedrich-Alexander-Universit{\"a}t Erlangen-N{\"u}rnberg, D-91058 Erlangen, Germany}
\affiliation{Physik-department, Technische Universit{\"a}t M{\"u}nchen, D-85748 Garching, Germany}
\affiliation{D{\'e}partement de physique nucl{\'e}aire et corpusculaire, Universit{\'e} de Gen{\`e}ve, CH-1211 Gen{\`e}ve, Switzerland}
\affiliation{Dept. of Physics and Astronomy, University of Gent, B-9000 Gent, Belgium}
\affiliation{Dept. of Physics and Astronomy, University of California, Irvine, CA 92697, USA}
\affiliation{Karlsruhe Institute of Technology, Institute for Astroparticle Physics, D-76021 Karlsruhe, Germany }
\affiliation{Karlsruhe Institute of Technology, Institute of Experimental Particle Physics, D-76021 Karlsruhe, Germany }
\affiliation{Dept. of Physics, Engineering Physics, and Astronomy, Queen's University, Kingston, ON K7L 3N6, Canada}
\affiliation{Department of Physics {\&} Astronomy, University of Nevada, Las Vegas, NV, 89154, USA}
\affiliation{Nevada Center for Astrophysics, University of Nevada, Las Vegas, NV 89154, USA}
\affiliation{Dept. of Physics and Astronomy, University of Kansas, Lawrence, KS 66045, USA}
\affiliation{Centre for Cosmology, Particle Physics and Phenomenology - CP3, Universit{\'e} catholique de Louvain, Louvain-la-Neuve, Belgium}
\affiliation{Department of Physics, Mercer University, Macon, GA 31207-0001, USA}
\affiliation{Dept. of Astronomy, University of Wisconsin{\textendash}Madison, Madison, WI 53706, USA}
\affiliation{Dept. of Physics and Wisconsin IceCube Particle Astrophysics Center, University of Wisconsin{\textendash}Madison, Madison, WI 53706, USA}
\affiliation{Institute of Physics, University of Mainz, Staudinger Weg 7, D-55099 Mainz, Germany}
\affiliation{Department of Physics, Marquette University, Milwaukee, WI, 53201, USA}
\affiliation{Institut f{\"u}r Kernphysik, Westf{\"a}lische Wilhelms-Universit{\"a}t M{\"u}nster, D-48149 M{\"u}nster, Germany}
\affiliation{Bartol Research Institute and Dept. of Physics and Astronomy, University of Delaware, Newark, DE 19716, USA}
\affiliation{Dept. of Physics, Yale University, New Haven, CT 06520, USA}
\affiliation{Columbia Astrophysics and Nevis Laboratories, Columbia University, New York, NY 10027, USA}
\affiliation{Dept. of Physics, University of Oxford, Parks Road, Oxford OX1 3PU, UK}
\affiliation{Dipartimento di Fisica e Astronomia Galileo Galilei, Universit{\`a} Degli Studi di Padova, 35122 Padova PD, Italy}
\affiliation{Dept. of Physics, Drexel University, 3141 Chestnut Street, Philadelphia, PA 19104, USA}
\affiliation{Physics Department, South Dakota School of Mines and Technology, Rapid City, SD 57701, USA}
\affiliation{Dept. of Physics, University of Wisconsin, River Falls, WI 54022, USA}
\affiliation{Dept. of Physics and Astronomy, University of Rochester, Rochester, NY 14627, USA}
\affiliation{Department of Physics and Astronomy, University of Utah, Salt Lake City, UT 84112, USA}
\affiliation{Oskar Klein Centre and Dept. of Physics, Stockholm University, SE-10691 Stockholm, Sweden}
\affiliation{Dept. of Physics and Astronomy, Stony Brook University, Stony Brook, NY 11794-3800, USA}
\affiliation{Dept. of Physics, Sungkyunkwan University, Suwon 16419, Korea}
\affiliation{Institute of Physics, Academia Sinica, Taipei, 11529, Taiwan}
\affiliation{Dept. of Physics and Astronomy, University of Alabama, Tuscaloosa, AL 35487, USA}
\affiliation{Dept. of Astronomy and Astrophysics, Pennsylvania State University, University Park, PA 16802, USA}
\affiliation{Dept. of Physics, Pennsylvania State University, University Park, PA 16802, USA}
\affiliation{Dept. of Physics and Astronomy, Uppsala University, Box 516, S-75120 Uppsala, Sweden}
\affiliation{Dept. of Physics, University of Wuppertal, D-42119 Wuppertal, Germany}
\affiliation{Deutsches Elektronen-Synchrotron DESY, Platanenallee 6, 15738 Zeuthen, Germany }

\author[0000-0001-6141-4205]{R. Abbasi}
\affiliation{Department of Physics, Loyola University Chicago, Chicago, IL 60660, USA}

\author[0000-0001-8952-588X]{M. Ackermann}
\affiliation{Deutsches Elektronen-Synchrotron DESY, Platanenallee 6, 15738 Zeuthen, Germany }

\author{J. Adams}
\affiliation{Dept. of Physics and Astronomy, University of Canterbury, Private Bag 4800, Christchurch, New Zealand}

\author[0000-0002-9714-8866]{S. K. Agarwalla}
\altaffiliation{also at Institute of Physics, Sachivalaya Marg, Sainik School Post, Bhubaneswar 751005, India}
\affiliation{Dept. of Physics and Wisconsin IceCube Particle Astrophysics Center, University of Wisconsin{\textendash}Madison, Madison, WI 53706, USA}

\author[0000-0003-2252-9514]{J. A. Aguilar}
\affiliation{Universit{\'e} Libre de Bruxelles, Science Faculty CP230, B-1050 Brussels, Belgium}

\author[0000-0003-0709-5631]{M. Ahlers}
\affiliation{Niels Bohr Institute, University of Copenhagen, DK-2100 Copenhagen, Denmark}

\author[0000-0002-9534-9189]{J.M. Alameddine}
\affiliation{Dept. of Physics, TU Dortmund University, D-44221 Dortmund, Germany}

\author{N. M. Amin}
\affiliation{Bartol Research Institute and Dept. of Physics and Astronomy, University of Delaware, Newark, DE 19716, USA}

\author{K. Andeen}
\affiliation{Department of Physics, Marquette University, Milwaukee, WI, 53201, USA}

\author[0000-0003-2039-4724]{G. Anton}
\affiliation{Erlangen Centre for Astroparticle Physics, Friedrich-Alexander-Universit{\"a}t Erlangen-N{\"u}rnberg, D-91058 Erlangen, Germany}

% ...

\date{\today}

\collaboration{406}{IceCube Collaboration}

\begin{abstract}
The IceCube Neutrino Observatory has been continuously taking data to search for $\mathcal{O}(0.5-10)$\,s long neutrino bursts since 2007.
Even if a Galactic core-collapse supernova is optically obscured or collapses to a black hole instead of exploding, it will be detectable via the $\mathcal{O}(10)$\,MeV neutrino burst emitted during the collapse. 
%A detection of an $\mathcal{O}(10)$ MeV neutrino burst would lead to a detection of galactic core collapse supernova (CCSN) even if those that are optically obscured or fail to explode due to the creation of a black hole. 
We discuss a search for such events covering the time between April 17, 2008 and December 31, 2019. Considering the average data taking and analysis uptime of 91.7\,\% after all selection cuts, this is equivalent to 10.735 years of continuous data taking. In order to test the most conservative neutrino production scenario, the selection cuts were optimized for a model based on a 8.8 solar mass progenitor collapsing to an O-Ne-Mg core. Conservative assumptions on the effects of neutrino oscillations in the exploding star were made. The final selection cut was set to ensure that the probability to detect such a supernova within the Milky Way exceeds 99\,\%.  
No such neutrino burst was found in the data after performing a blind analysis. Hence, a 90\,\% C.L. upper limit on the rate of core-collapse supernovae out to distances of $\approx$ 25\,kpc was determined to be 0.23/yr.  For the more distant Magellanic Clouds, only high neutrino luminosity supernovae will be detectable by IceCube, unless external information on the burst time is available. We determined a model-independent limit by parameterizing the dependence on the neutrino luminosity and the energy spectrum.
\end{abstract}

\keywords{supernova --- neutrino --- galactic}
 
 \section{Introduction}\label{s:introduction}
Stars with masses larger than 
$\approx 8\,\mathrm{M}_\odot$ end their lives with the gravitational collapse of their core, followed by neutrino emission over a time scale of about 10\,s and a shock-driven luminous explosion called a supernova. The expected volumetric rate of core-collapse supernovae (CCSNe) in the universe today is $ R_\mathrm{SN}\approx 10^{-4}/\mathrm{yr}/\mathrm{Mpc}^3$. 
By compiling results obtained with various methods, the rate of stellar collapses in the Milky Way, including those obscured in the optical waveband, was estimated  to be $1.63 \pm 0.46$ per century~\citep{Rozwadowska:2020nab}. This corresponds to a mean time between CCSNe of $T=61^{+24}_{-14}$ years. 

\begin{figure}[ht]
\centering
\includegraphics[angle=0,width=0.48\textwidth]{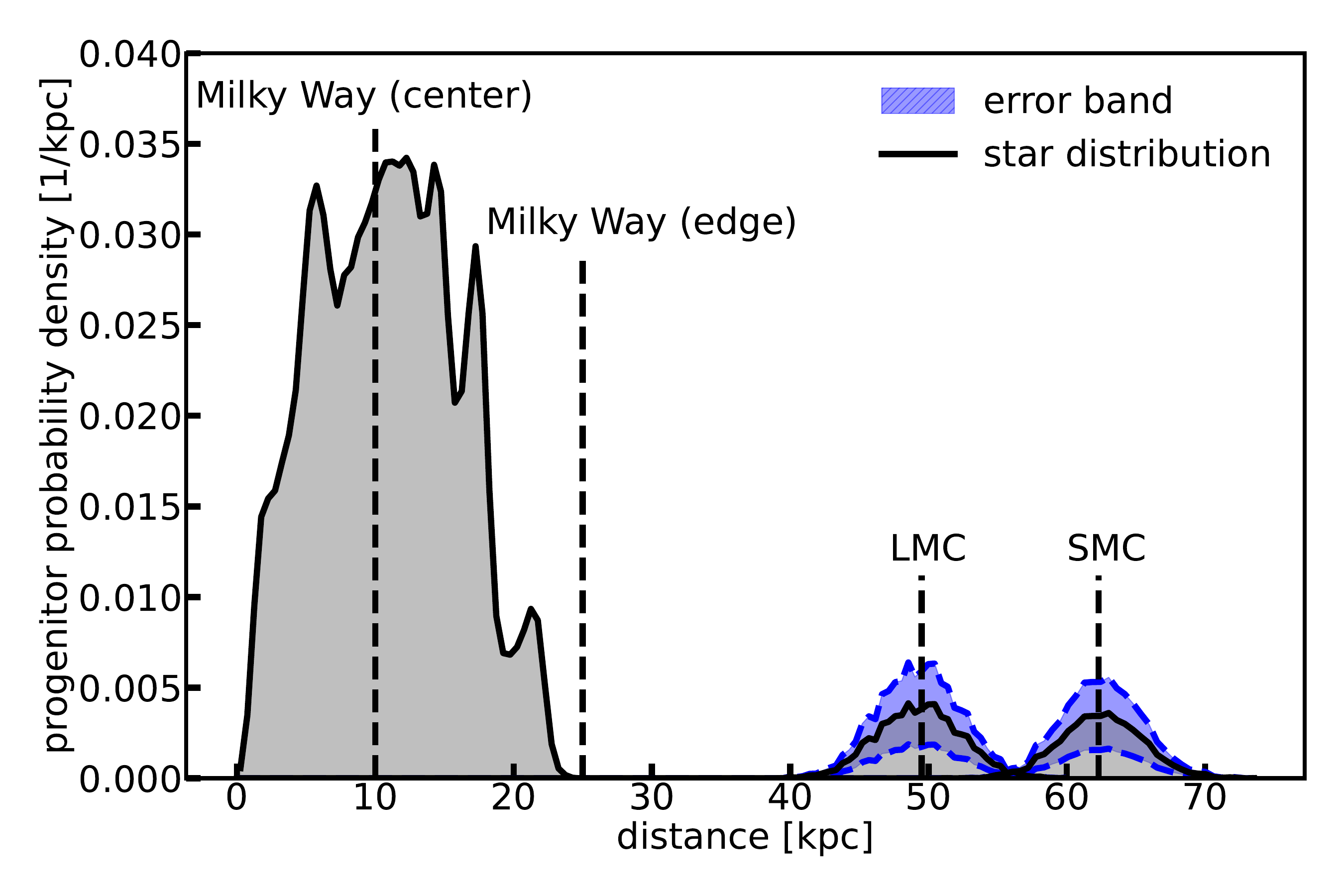}
\caption{\small Probability density distribution of progenitor distances in the Milky Way assumed in this paper. The spikes reflect the higher star density in the spiral arms. Also shown are additional relative contributions from the SMC and LMC with scale uncertainty estimates.The uncertainties do not enter the analysis.}\label{fig:distancedist}
\end{figure}

In spiral galaxies, observations indicate that CCSNe occur preferentially in the disc rather than the bulge component and therefore point to a relatively young progenitor population~\citep{10.1093/mnras/stv2853}.  
For this paper, we assume a Milky Way progenitor radial distribution model~\citep{Ahlers:2009ae} that takes into account the geometry of the spiral arms (see Fig.~\ref{fig:distancedist}). 

In principle, it is simpler to estimate the CCSN rate in the Large and Small Magellanic Clouds at $\approx 49.5$\,kpc and $\approx 62.8$\,kpc distance, respectively, because our view is not obscured by the dense bulge of the Milky Way. The rate can be estimated by counting the number of observed supernova remnants~\citep{2020pesr.book.....V} and the use of an isotope measure to distinguish between core collapse and Type Ia supernovae~\citep{Maggi:2015uga}. 
In spite of the relatively small number of visible stars in the Large and Small Magellanic Clouds compared to the Milky Way, both add estimated CCSN rates of $11\pm 6$\,\% to the Milky Way CCSN rate. 

The detection of neutrinos from CCSNe is important as they
reveal the conditions in the core region of the star at the time of the collapse. Neutrino experiments will also tell astronomers when and where they should point their telescopes, with several hours lead time~\citep{Super-Kamiokande:2016kji,Brdar_2018,PhysRevD.100.103005,refId0}. 

The fraction of supernovae that will be missed in optical observations depends on how regularly the complete sky is monitored by astronomers and automated systems and on whether the supernova is obscured by dust. Infrared photons can penetrate dust in the inner region of the galaxy. Assuming an optimistic model of dust extinction and all-sky coverage by optical telescopes, 96\,\% of the supernovae in the Milky Way should be observable in the optical~\citep{Adams:2013ana}.  For other dust models, this fraction may be lower, although dust extinction can be mitigated by automatic wide-angle infrared transient surveys~\citep{2019NatAs...3..109M}. Predictions have also been obtained for the more distant Magellanic Clouds~\citep{Adams:2013ana}.

Massive stars may create a black hole that consumes the nascent supernova before the massive explosion. This is indicated by an observed deficit of  supernova progenitors between 18 to 25\,$\mathrm{M}_\odot$~\citep{Kochanek:2008mp} and the merging of $> 20\,\mathrm{M}_\odot$ black holes observed by Advanced LIGO~\citep{LIGOScientific:2016aoc} which were likely formed from failed supernovae. The fraction of supernovae that end up in black holes is not well known, however.
A data-driven way to estimate the fraction is an optical search for progenitors that "suddenly disappear". 
An eleven year optical search~\citep{Neustadt:2021jjt} identified one clearly detected failed supernova candidate, corresponding to a 90\,\% confidence interval of $0.04 \le f \le 0.39$ for the fraction $f$ of core-collapses, resulting in failed or aborted supernovae.
Failed supernovae would be identifiable by the fast drop of the neutrino emission and a longer lasting echo of higher energy neutrinos that may be observable by IceCube~\citep{Gullin:2021hfv}.

It has been more than three decades since the first and only supernova has been observed by neutrino detection. 
On February 23, 1987, a burst of neutrinos with energies of a few tens of MeV emitted by
the supernova SN1987A was recorded simultaneously by the
Baksan~\citep{Alekseev:1987ej}, IMB~\citep{Bionta:1987qt}, and Kamiokande-II~\citep{Kamiokande-II:1987idp,Hirata:1988ad} detectors, a few hours before its optical counterpart was discovered. With just 24 neutrinos collected, important limits on the mass of the $\overline{\nu}_\mathrm{e}$, its lifetime, its magnetic moment and the number of leptonic flavors were derived~\citep{Kotake:2005zn}. 

The small observed number of events can be explained by the source distance ($\approx 50$\,kpc) and the limited active volumes of the detectors operational in 1987. As will be explained below, IceCube will register $\mathcal{O}(10^5 - 10^6)$  photons from interacting neutrinos if a supernova explodes at the Galactic Center. Such a high statistical accuracy will permit a study of detailed features of the neutrino emission that carries important information about the explosion dynamics and neutrino properties.  

Several neutrino detectors have searched for nearby CCSNe in the last decades, e.g. water Cherenkov detectors~\citep{AMANDA:2001htp,Super-Kamiokande:2007zsl,IceCube:2011cwc,KM3NeT:2021oaa},
scintillator detectors~\citep{MACRO:2004fzv,Novoseltsev:2019gdt,LVD:2014uzr,SNO:2010noh,Rumleskie:2020iip,KamLAND:2022sqb,NOvA:2021zhv},
lead-based detectors~\citep{Gallo_Rosso_2021}, as well as liquid noble gas neutrino~\citep{Abi_2021} and dark matter detectors~\citep{Lang_2016}.

The most stringent experimental limits have been published by the Baksan Collaboration~\citep{Novoseltsev:2022lmd} over the period June 30, 1980 to June 30, 2021 with a livetime of 35.5  years. In the absence of a positive observation, they quote a rate of < 6.5 CCSNe per century  within 20\,kpc at 90\,\% C.L.  
The LVD Collaboration quotes a rate of < 8 CCSNe per century  within 25\,kpc at 90\,\% C.L. over the period 1992 to Jan 4, 2021~\citep{Vigorito:2021sgy}.

Both LVD and Baksan determined their limits referencing phenomenological models that have been parameterized to fit the 1987A observation, and thus correspond to massive  $\mathcal{O}(20)\, \mathrm{M}_\odot$ progenitors. 
No neutrino oscillations were assumed by the Baksan Collaboration, whereas a normal neutrino hierarchy and MSW oscillations were taken into account in the LVD analysis. 

The possibility to monitor supernovae in our Galaxy with high-energy neutrino telescopes was first pointed out by~\cite{Pryor:1987tz} and ~\cite{Halzen:1995ex}. 
A first search~\citep{AMANDA:2001htp} based on 215 days of data taken in 1997 and 1998 with the still incomplete AMANDA detector demonstrated the feasibility of the approach. Since 2009, IceCube has been sending real-time messages to the Supernova Early Warning System SNEWS~\citep{Antonioli:2004zb,SNEWS:2020tbu}. 

While the observation time of IceCube is shorter than that of Baksan and LVD, the large volume of IceCube provides sensitivity
to a wide variety of models, ranging from the lightest CCSNe to heavy progenitors that end up in a black hole. 

Following a description of supernova phenomenology and neutrino production (section~\ref{s:theory}), we briefly discuss the detection principle, data cleaning, statistics, and simulation (sections~\ref{s:detector}-~\ref{s:simulation}),  before we summarize the results (section~\ref{s:Iresults}) and conclude (section~\ref{s:conclusion}). 

\section{Supernovae and Neutrinos}\label{s:theory}
Neutrinos play a crucial role at all stages of the collapse of massive stars. In the initial phase of the collapse, the release of electron neutrinos by converting protons to neutrons accelerates the infall by removing the electron degeneracy pressure ("deleptonization phase"). The continuous accretion of outer layers of the progenitor star ("accretion phase") will eventually lead to the collapse and the formation of a dense and compact neutron star or a black hole. Matter bounces off this core ("core bounce") and emits a shock wave. The absorption of electron neutrinos and antineutrinos in the material surrounding the neutron star invigorates the shock so that the star is blown apart. Due to their small interaction cross section, only neutrinos and antineutrinos of all flavors carry away the gravitational binding energy of the compact and dense remnant ("cooling phase"). Subtleties in neutrino interactions, oscillations, and transport play a surprisingly large role. 
For a compilation on many aspects of supernova research see~\cite{2017hsn..book.....A}. 

The  supernova core is sufficiently hot and dense to host a thermal population of neutrinos of all species that diffuse out and eventually reach the Earth. The neutrino thermal energy spectrum is expected to peak between 10 to 20\,MeV, with $\nu_\mathrm{e}$ and $\overline{\nu}_\mathrm{e}$ carrying lower energies as they are more strongly coupled to matter and evaporate later than $\mu$ and $\tau$ neutrinos. Neutrinos carry away 99\,\% of the gravitational binding energy released in the collapse, typically $\approx 3\times 10^{46}$\,J, roughly equally distributed between the six neutrino and antineutrino species. 

Flavor mixing effects can change the expected neutrino rates  as well as the energy spectra compared to the original time-dependent $\nu_e$ flux, $F_{\nu_\mathrm{e}}^0$, and $\overline{\nu}_\mathrm{e}$ flux, $F_{\mathrm{\overline{\nu}_\mathrm{e}}}^0$.
Deep inside the core, where the neutrino mean free path is comparable to the size of the proto-neutron star, flavor mixing may be ignored. 
Further away from the core, but within  $\approx 200$ km from the center of the star, the density of neutrinos exceed that of electrons and the coherent scattering  of neutrinos on each other can no longer be neglected~\citep{PhysRevD.88.023008}.  Such collective effects 
after the core bounce may lead to complex energy and time-dependent neutrino flavor conversions and the swapping of electron neutrinos with muon and tau neutrinos. At larger radii, the neutrino flavor conversion is driven by coherent scattering on electrons. Resonant enhancements for flavor conversion occur at densities around $10^6$\,kg/m$^3$ and $10^4$\,kg/m$^3$. Already in the vicinity of the supernova, the coherence of the mass eigenstates is lost, leading to wave packet separations at Earth of many meters~\citep{Kersten:2015kio}. 

The second and third generation neutrinos $ \nu_\mathrm{x} := \nu_\mu, \overline{\nu}_\mu, \nu_\tau, \overline{\nu}_\tau$ are mostly produced in the cooling phase and their fluxes are roughly equal. 
The mixing of the $\nu$ and $\overline{\nu}$ fluxes is then given by
\begin{eqnarray}
F_{\nu_\mathrm{e}} &=& p\cdot F_{\nu_\mathrm{e}}^0+(1-p)\cdot F_{\nu_\mathrm{x}}^0 \\
F_{\overline{\nu}_\mathrm{e}} &=& \overline{p}\cdot F_{\overline{\nu}_\mathrm{e}}^0+(1-\overline{p})\cdot F_{\overline{\nu}_\mathrm{x}}^0\quad .
\end{eqnarray}

The probabilities $p$, $\overline{p}$ depend on the $\theta_{12}$ mixing angle, collective effects, state transitions probabilities and matter densities~\citep{rivistareview}.
If one neglects collective effects and assumes a static supernova matter profile with adiabatic state transitions, one obtains the simple relations 
\begin{eqnarray}
p &=& 0\;;\;\; \overline{p}=\cos^2\theta_{12} \quad (\text{NH)}\label{eqNH}\\ 
\overline{p}&=& 0\;;\;\; p=\sin^2\theta_{12}\;\quad \text{(IH)}\label{eqIH}
\end{eqnarray} 
for the normal neutrino mass hierarchy (NH) and inverted hierarchy (IH). When testing the effect of oscillations on models, we will use equations~\ref{eqNH} and \ref{eqIH}, as well as the no-oscillation case. 

Oscillation effects can alter the detected signal as cross sections, fluxes, and energy spectra vary between flavors. While energies and fluxes are similar during the cooling phase, substantial differences in the early phase of neutrino emission may strongly modify the time-dependent flux. As the CCSN neutrino-induced hit rate in IceCube roughly rises with $E_\nu^3$~\citep{IceCube:2011cwc}, it is particularly sensitive to oscillation-induced changes in the energy spectra. For each model we chose the oscillation scenario that leads to the lowest rate.

Finally, when neutrinos enter the material of the Earth, oscillations will occur that depend on the neutrino energy, the path length in matter and the material density. The resulting effect on the measured signal in IceCube will be discussed as part of the systematic uncertainties.

The initial phase of the neutrino emission is rather insensitive to the supernova progenitor mass. The total energy release in neutrinos depends directly on the mass of the neutron star~\citep{Lattimer:2000nx} and only indirectly on the supernova progenitor mass. From 22 historic supernova remnants,~\cite{Diaz-Rodriguez:2021xja} find that the progenitor mass distribution is proportional to $M^a$, with $a = -2.61^{+1.05}_{-1.18}$ with a minimal progenitor mass of $8.60^{+0.37}_{-0.41}$\,$\mathrm{M}_\odot$. Assuming these values, a substantial fraction, 10\,\% to 40\,\%, of all supernova progenitors, would have masses below 20\,$\mathrm{M}_\odot$. 

Various theory groups have performed extensive simulations of the supernova neutrino emission with ever increasing level of detail ~(e.g. \cite{JankaReview,Burrows:2020qrp}). 
Public codes exist that provide
links between such simulations and simulations of neutrino detectors~\citep{Migenda:2021hnl,baxter2021snewpy} and to quickly test physics signatures in current and future detectors~\citep{snowglobes,Malmenbeck:2019giw}. While state-of-the-art calculations are performed in three-dimensional space, such calculations are very time consuming and only a few cover more than the first few hundred milliseconds until the explosion takes place (or not). For very low mass progenitors, and the early and late phases of neutrino emission, spherically symmetric simulations work reasonably well and will often lead to a neutrino-driven explosion in the simulation. 

For the lowest expected signal we chose a $\approx 9$\,s long simulation of a supernova from a 8.8$\,\mathrm{M}_\odot$ progenitor star that is triggered by electron-capture reactions and forms an O-Ne-Mg core~\citep{Hudepohl:2009tyy}, henceforth referred to as the ``H\"udepohl model''. With a total emitted energy of $1.7\,\times\, 10^{39}$\,J and a low mean neutrino energy of $\langle E_{\overline{\nu}_\mathrm{e}}\rangle \approx 12.9$\,MeV, it represents a conservative lower limit for a supernova search.

The second choice is a 19\,$\mathrm{M}_\odot$ progenitor whose collapse was modeled in three dimensions up to 1.756\,s after the core bounce~\citep{Bollig:2020phc}. Using an adaptive procedure, this simulation was stitched to a spherically symmetric simulation that continued into the cooling phase.
The third model starts from a 
27\,$\mathrm{M}_\odot$ model~\citep{Burrows:2020qrp} and follows the collapse and explosion with a cylindrically symmetric calculation that extends up to ~4.5\,s post-bounce.

On the high mass side, the gravitational collapse of stars exceeding $\mathcal{O}(25)\,\mathrm{M}_\odot$  will lead to a partial stellar explosion, while stars exceeding $\mathcal{O} (50)\,\mathrm{M}_\odot$ are not expected to explode at all~\citep{Smartt_2015,O’Connor2017}. In both cases, a black hole will develop $\mathcal{O}$(1)\, s after core bounce. At this point, the neutrino emission vanishes abruptly  in non-rotating systems. For the analysis presented in this paper, we select a model assuming a 40\,$\mathrm{M}_\odot$ progenitor and a hard equation of state~\citep{SHEN1998435} for the neutron star~\citep{Sumiyoshi:2007pp}.  This one-dimensional simulation of a non-rotating star ends in a black hole after $\approx$ 1.3\,s. The model was also used in LVD's result~\citep{Vigorito:2021sgy}.

The time series of the neutrino emission differs substantially between the models. While it has been shown for one-dimensional simulations that most of the codes agree between various groups within 5\,\%~\citep{OConnor:2018sti}, there are substantial differences when extending the calculations to more independent dimensions. 

In the figures,  we will refer to the models discussed in this section with the short forms "$8.8\,\mathrm{M}_\odot$", "$19\,\mathrm{M}_\odot$", "$27\,\mathrm{M}_\odot$", and  "$40\,\mathrm{M}_\odot$". The two low (high) progenitor mass models yield the lowest rate in the no-oscillation (inverted hierarchy MSW) case.

To summarize, neutrinos are crucial during all stages of the stellar collapse and the explosion. They are, besides gravitational waves, the only means to obtain immediate information from the central regions of a dying star. 
\vspace{1.5cm}
\section{The detection principle of IceCube}\label{s:detector}
The IceCube Neutrino Observatory is a cubic-kilometer  
Cherenkov detector  installed in the ice at the geographic South Pole~\citep{IceCube:2016zyt}  between depths of 1450\,m and 
2450\,m. The detector was
constructed from January 28, 2005 to December 18, 2010 by drilling  holes into the Antarctic ice
sheet in a hexagonal grid layout. Eighty-six cables (known as ``strings''), instrumented with digital optical modules (DOMs) containing 10 inch hemispherical Hamamatsu
R7081 photomultiplier tube~\citep{IceCube:2010dpc}, were lowered into the ice. Strings are spaced
about 125\,m apart and each contains 60 DOMs vertically
separated by 17\,m. Six central strings, with smaller string and sensor separations and equipped with $\approx$ 35\,\% higher quantum efficiency photosensors (HQE), are used in the DeepCore sub-array~\citep{IceCube:2011ucd}. The data used in this paper include the construction period where 2400, 3540, and 4740 DOMs became available to the data analysis during the first half of 2008, 2009 and 2010, respectively. The final detector consists of 5160 DOMs, of which  98.4\,\% are available to the analysis~\citep{IceCube:2016zyt}.

IceCube was designed to detect neutrinos with TeV energies and above. However, the neutrinos expected from CCSNe typically carry only $\mathcal{O}(10)$ MeV energies and only about 0.2\,\% of interactions within the detector volume will lead to at least one detected Cherenkov photon produced by secondary particles. Still, with a sufficient number of low-energy neutrinos interacting in the detector volume, a Galactic CCSN will produce a detectable correlated rise in the hit rate of all DOMs. Details of the detection method, the data acquisition system, and the physics capabilities can be found in~\cite{IceCube:2011cwc}.  

A dedicated pulse counter-based data acquisition is used to search for signatures of CCSNe. The search algorithm is based on count rates of individual DOMs stored in 1.67\,ms time bins, which are downsampled to 500\,ms intervals to perform various statistical analyses. An artificial deadtime of $\tau=247.5$\,$\mu$s was introduced to suppress time-correlated supra-Poissonian photomultiplier pulses at low temperatures that are  most probably due to a temperature dependent radiative dissipation of energy deposited by radioactive decays in the glass~\citep{Meyer_2010,HeeremanvonZuydtwyck:2015mbs}.
This deadtime leads to an inefficiency that can be parameterized by $\epsilon_\mathrm{dead time}\approx 0.87/(1+r_\mathrm{SN}\cdot \tau)$, where  $r_{SN}$ denotes the excess rate per optical module from a CCSN~\citep{IceCube:2011cwc}. The resulting DOM background rate is below 300 Hz (see Fig.~\ref{fig:timeseries}).

The stability of the DOM background rates is crucial for IceCube's sensitivity to detect supernovae. By using automatic online procedures, faulty modules are excluded while acquiring the data.  
In the final 86-string configuration, $\approx$ 1.4\,\% of the 5160 modules were permanently excluded from the analysis and only modules with dynamically calculated background rates below 10\,kHz were accepted.  
Operational modules were removed from the analysis if they exhibited either a variance  much larger than the Poissonian expectation or a high skewness~\citep{IceCube:2011cwc}. Typically, only one or two DOMs are affected by the real time quality selection. 
In the very rare case where the number of qualified modules drops below a threshold of 100, the corresponding time periods were discarded as a safeguard to prevent sending false alerts to SNEWS. 

By buffering the full photomultiplier raw data stream that is stored around candidate neutrino burst or external alerts, additional information can be retrieved~\citep{HeeremanvonZuydtwyck:2015mbs,IceCube:2016zyt}. For example, the average $\overline{\nu}_\mathrm{e}$ energy can be estimated from rare coincidences between adjacent DOMs~\citep{Fritz:2021btf} and the precision of the burst onset time can be improved.

Inverse beta decay, $\overline{\nu}_\mathrm{e}+\mathrm{p}\rightarrow \mathrm{n} + \mathrm{e}^+ $, dominates the interaction in water or ice
~\citep{IceCube:2011cwc}.
The signal hit rate per DOM for the inverse beta decay is given by
\begin{eqnarray}
	R(t) & = & \epsilon_\mathrm{dead time}\frac{n_\mathrm{target} \,\mathcal{L}_\mathrm{SN}^\mathrm{\nu}(t)}{4\pi d^2\langle E_\mathrm{\nu}(t)\rangle} 
	 \int_0^{\infty} \, dE_\mathrm{e^+} \!\! \int_0^{\infty}  \, dE_\mathrm{\nu} \nonumber \\
	&\times\, & \frac{d\sigma}{dE_\mathrm{e^+}}(E_\mathrm{e^+},E_\mathrm{\nu}) \,V_\mathrm{e^+}^\mathrm{eff} \,f(E_{\nu},\langle E_\mathrm{\nu}\rangle, \alpha_\nu,t) \, , 
	  \label{eq:rate}
\end{eqnarray}
where $n_\mathrm{target}$ is the density of proton targets in ice, $d$ is the distance to the supernova, $\mathcal{L}_\mathrm{SN}^\nu(t)$
its neutrino luminosity, and $f(E_\mathrm{}{\nu},\langle E_\mathrm{\nu}\rangle, \alpha_\mathrm{\nu}, t)$ is the normalized $E_\mathrm{\nu}$ distribution depending on a shape parameter $\alpha_\nu$ and on the average neutrino energy $\langle E_\mathrm{\nu}\rangle $. $E_\mathrm{e^+}$ denotes the energy of the positron emerging from the neutrino interaction. The effective volume for a single positron, 
$V^\mathrm{eff}_\mathrm{e^+}$,
strongly varies with the photon absorption in the ice but shows little dependence on photon scattering. It is also directly proportional to the positron track length and thus to the positron energy~\citep{IceCube:2011cwc}.

\section{Data preparation}\label{s:data}
The data used in this analysis cover the time between April 17, 2008, 03:43:54.16 UTC and December 31, 2019, 19:12:24.97 UTC (11.707 years) and were taken with four detector configurations of 40, 59, 79 and 86 strings, respectively.  The data taking efficiency improved with time from 96\,\% in 2009 to 99.7\,\% in 2017. The average livetime was 98.4\,\% during this period. 

The analysis
requires that the detector works faultlessly in each of the $\approx 700$ million half-second time intervals studied. Therefore, additional measures are required to clean the data. Short runs (< 20 min), runs taken with calibration light sources, with an incomplete detector configuration, or containing data taking errors, were discarded. The total number of contributing DOMs was required to be larger than a minimum number, for example, 5060 DOMs out of a total of 5160 DOMs in case of the final IceCube detector configuration. We also required that the data acquisition for reconstructing muon tracks was working perfectly and that there was no known electromagnetic interference from external sources, such as radar surveys of the experimental site at the South Pole. 

After rejecting such problematic data, the uptime available to the analysis ranged between 86.6\,\% and 96.8\,\%, with an average value of 91.7\,\%. The selected clean data, joined together, would correspond to 10.735 years of continuous data taking.

Atmospheric muons constitute a background to the search for CCSNe even though their energy when entering the ice sheet needs to be above $\approx$ 550 GeV to trigger the IceCube 8-fold majority trigger~\citep{Kelley:2015ncf}.
Hence, hits from muon tracks that fail the trigger requirements are mostly found in the upper detector layers or clip the corners of the detector. Due to atmospheric density changes that are correlated to the air temperature, the muon rate shows a seasonal dependence and short term variations. 

\begin{figure}[ht]
\centering
\includegraphics[angle=0,width=0.48\textwidth]{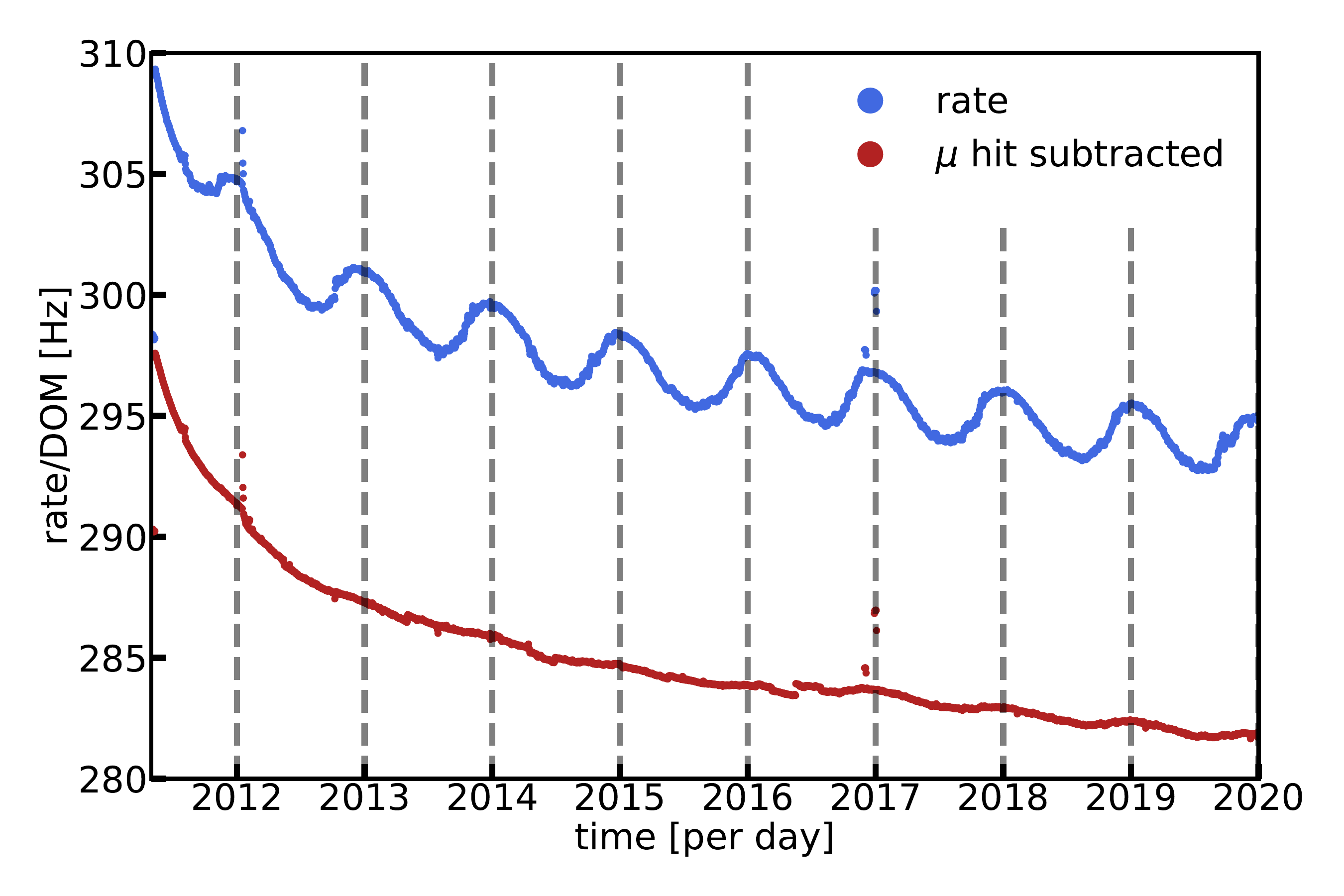}
\caption{\small The blue dots show the average DOM photoelectron count rate for IceCube DOMs with standard efficiency photomultipliers. The red curve shows the rate after hits associated with muon tracks have been removed. Note that the data still show rare artifacts due to sporadic rate increases of individual DOMs. This effect is addressed with the test statistic discussed in section~\ref{s:teststatistics}. }\label{fig:timeseries}
\end{figure}

Because dust layers absorb light and many muons range out, the  atmospheric muon-induced rate is depth-dependent 
and adds  3\,Hz to 30\,Hz to the $\sim 280$\,Hz  optical module background rate. This can be seen in Fig.~\ref{fig:timeseries}, where the averaged count rate per DOM is shown as a function of time (blue points). The contribution from muons varies with the season. The red points show the rates after hits from identified muons were removed; the seasonal effect due to atmospheric muons is strongly reduced. The $\approx 5$\,\% reduction in rate is thought to be due to relieved stress in the refrozen ice near each DOM that decreases the effect of triboluminescence.  It has been verified that the effect is not due to PMT aging~\citep{Aartsen_2020}. 
%\vspace{1cm}
\section{Test statistic for the supernova search}\label{s:teststatistics}
The test statistic used to search for Galactic supernovae with IceCube is the significance proxy 
\begin{equation}
\xi = \frac{\Delta\mu}{\sigma_{\Delta\mu}}\quad ,
\end{equation}
where 
\begin{equation}
	\Delta\mu = \sigma_{\Delta\mu}^2 \sum_{i=1}^{N_\mathrm{DOM}} \, \frac{\epsilon_i\,(r_i - \langle r_i\rangle)}{\langle\sigma_i\rangle^2}
\end{equation}
is the most likely collective rate deviation of all  $N_\mathrm{DOM}$ rates $r_i$ from their running average. The average of the rates, $\langle r_i\rangle$, and the corresponding standard deviations $\langle\sigma_i\rangle$ are calculated from sliding 285\,s  
time intervals before and after the central investigated time interval of 29.5\,s duration. The factors $\epsilon_i$ account for quantum efficiency differences of the DOMs.
Note that $\Delta\mu$ has the structure of a weighted average sum. The squared uncertainty on   $\Delta\mu$,  
\begin{equation}
	\sigma_{\Delta\mu}^2 = \left(\sum_{i=1}^{N_\mathrm{DOM}} \, \frac{{\epsilon_i}^2}{\langle\sigma_i\rangle^2}\right)^{-1} \quad ,
\end{equation}
is calculated from the data and thus accounts  for non-Poissonian behavior in the background rates. In purely uncorrelated Poissonian processes, the significance should be centered at zero with unit width. 

\begin{figure}[ht]
\centering
\includegraphics[angle=0,width=0.52\textwidth]{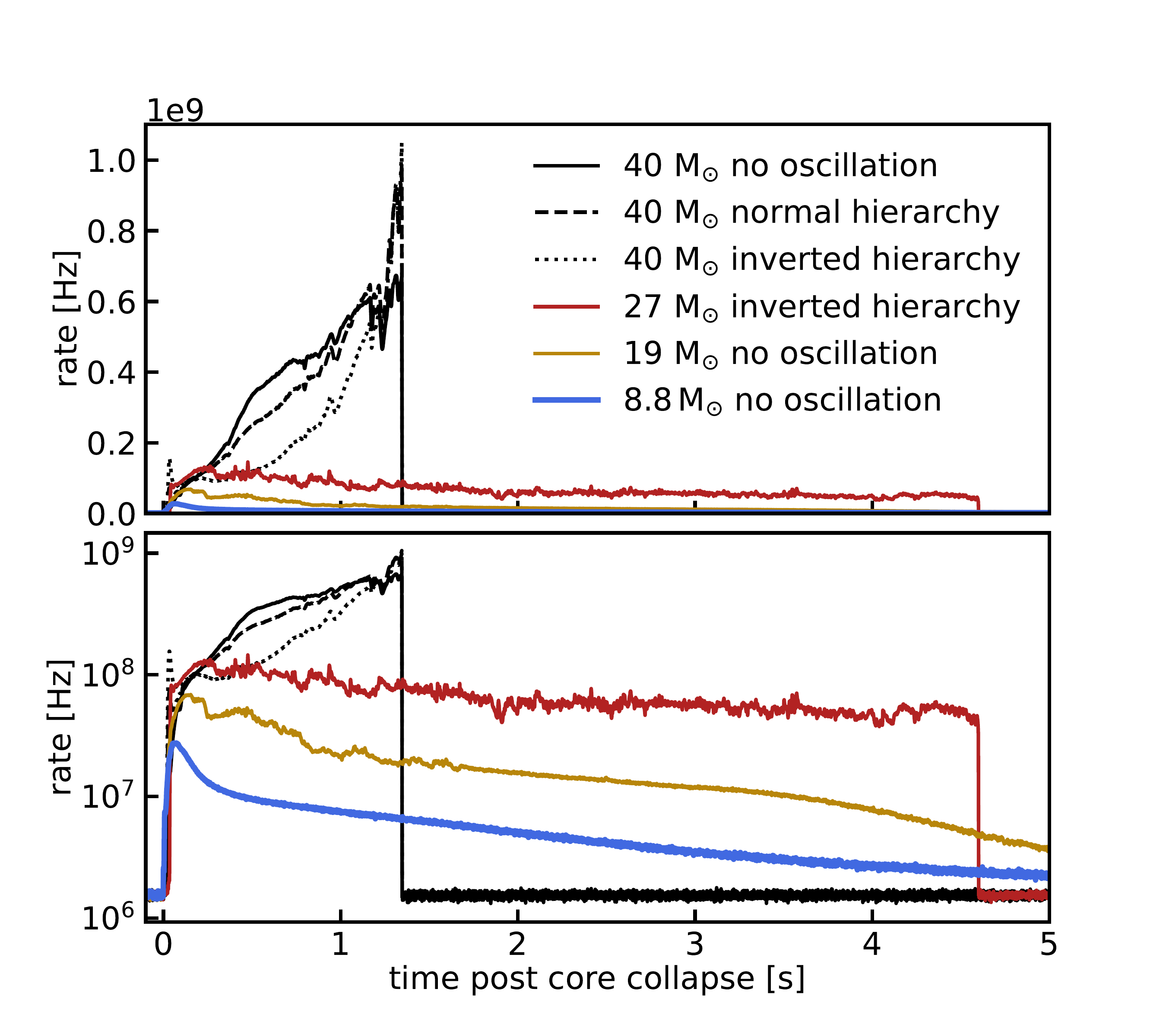}
\caption{The number of hits in IceCube is shown for the four investigated models assuming a short supernova distance of 1\,kpc,  both in linear and log scale. The observed baseline is due to the background rate. Earth oscillation effects have not been included. Note that the signal rate is roughly proportional to $E_\nu^3$. Hence, models with higher neutrino energies are much more prominently seen. For the example of black hole formation, the effect of MSW oscillations is demonstrated. The effects are much smaller for the low mass Huedepohl model (blue), used as a conservative benchmark in the analysis, where the no oscillation case yields the lowest hit rate. }\label{fig:models}
\end{figure}

The calculation in the data acquisition was done in consecutive, non-overlapping 500\,ms wide time intervals as well as in sliding 1.5, 4, 10\,s time intervals overlapping by 500\,ms.
The sliding window approach introduces correlations and -- picking the highest significance -- distorts the Gaussian shape of the distribution by adding a high significance tail. We chose the 
1.5\,s time binning for all analyses as a conservative compromise among the models that were tested. As shown in Fig.~\ref{fig:models}, it covers the accretion phase with high neutrino intensity well.

The effect of muons on the significance proxy $\xi$ is much more pronounced than in the summed hits because muons create space and time-correlated hits. Therefore the optical sensor rates are no longer statistically independent  and the central limit theorem is only partially fulfilled. 
 
 In fact, the vast majority of false positive alerts are due to a statistical clustering of atmospheric 
 muon-induced hits: the rate of false alerts is cut by almost three orders of magnitude after removing hits associated with atmospheric muons.
 
 In order to properly account for DOM rate variations, we correct for the muon contribution by a decorrelation method, which has been applied before in~\cite{IceCube:2019acm}.
We define a muon significance proxy by the relation
\begin{eqnarray}
\xi_\mu & = &\frac{R_\mu^\mathrm{hit}-\langle R_\mu^\mathrm{hit}\rangle }{\sigma(R_\mu^\mathrm{hit})}\quad,
\end{eqnarray}
where all quantities are calculated on a per run basis.  $R_\mu^{\text{hit}}$, the sum over all hits associated to a muon track, is taken as a measure of the atmospheric muon intensity.

\begin{figure}[ht]
\centering
\includegraphics[angle=0,width=0.48\textwidth]{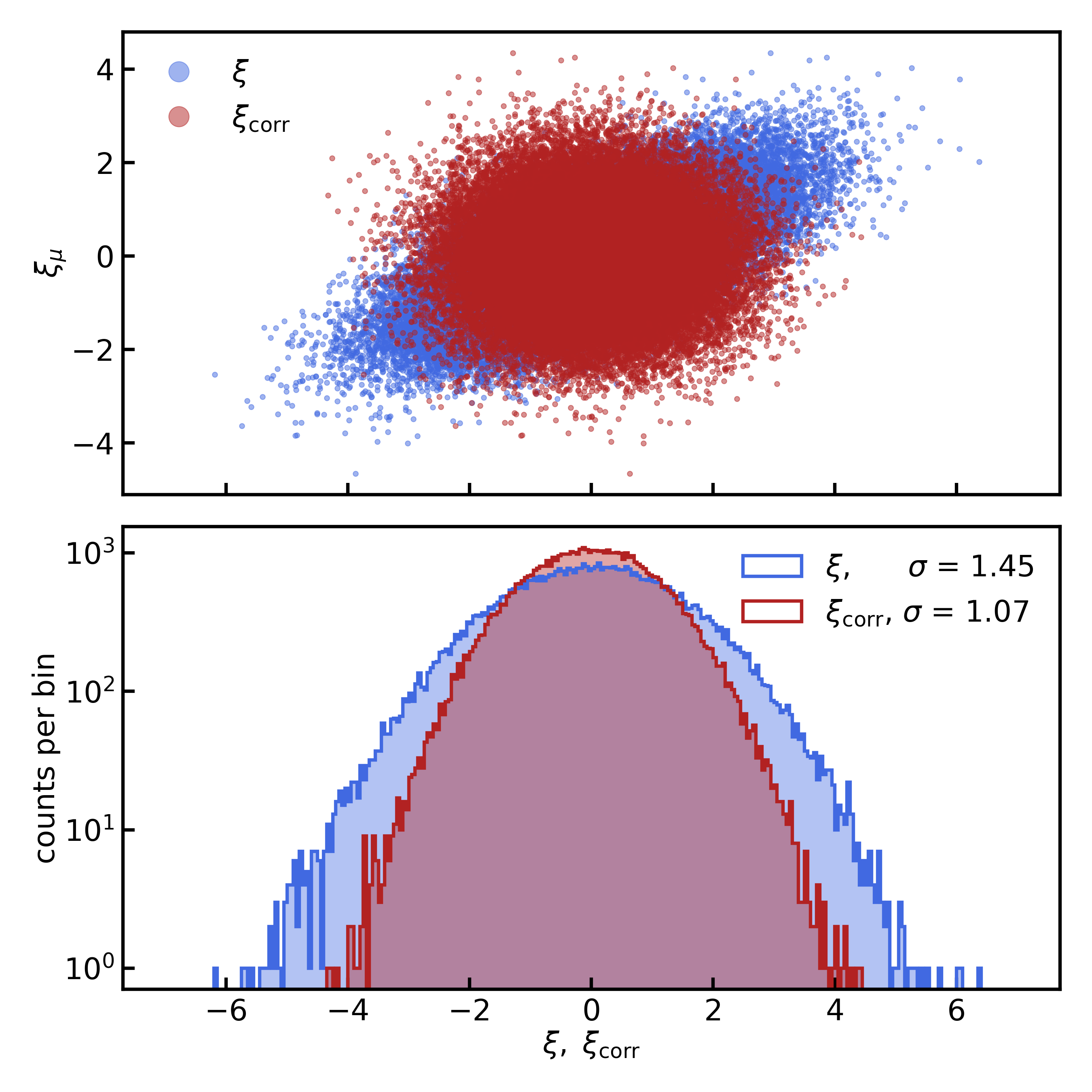}
\caption{\small Upper plot: correlation of the atmospheric muon significance proxy  $\xi_{\mu}$  with the significance proxies $\xi$ and $\xi_\mathrm{corr}$. The application of the decorrelation discussed in the text leads to the red distribution. Lower plot: the projections demonstrate the sharpening of the significance distribution after decorrelation.
}\label{fig:correlation}
\end{figure}

A linear function is fit to the correlation between the supernova significance proxy $\xi$ and $\xi_\mu$
(see Fig.~\ref{fig:correlation} for an example 8 hour run). A corrected significance proxy $\xi_\mathrm{corr} = \xi - b \cdot R_\mu^{\text{hit}} - a$ is calculated from the resulting offset $a$ and slope $b$. 

\begin{figure}
\includegraphics[angle=0,width=0.48\textwidth]{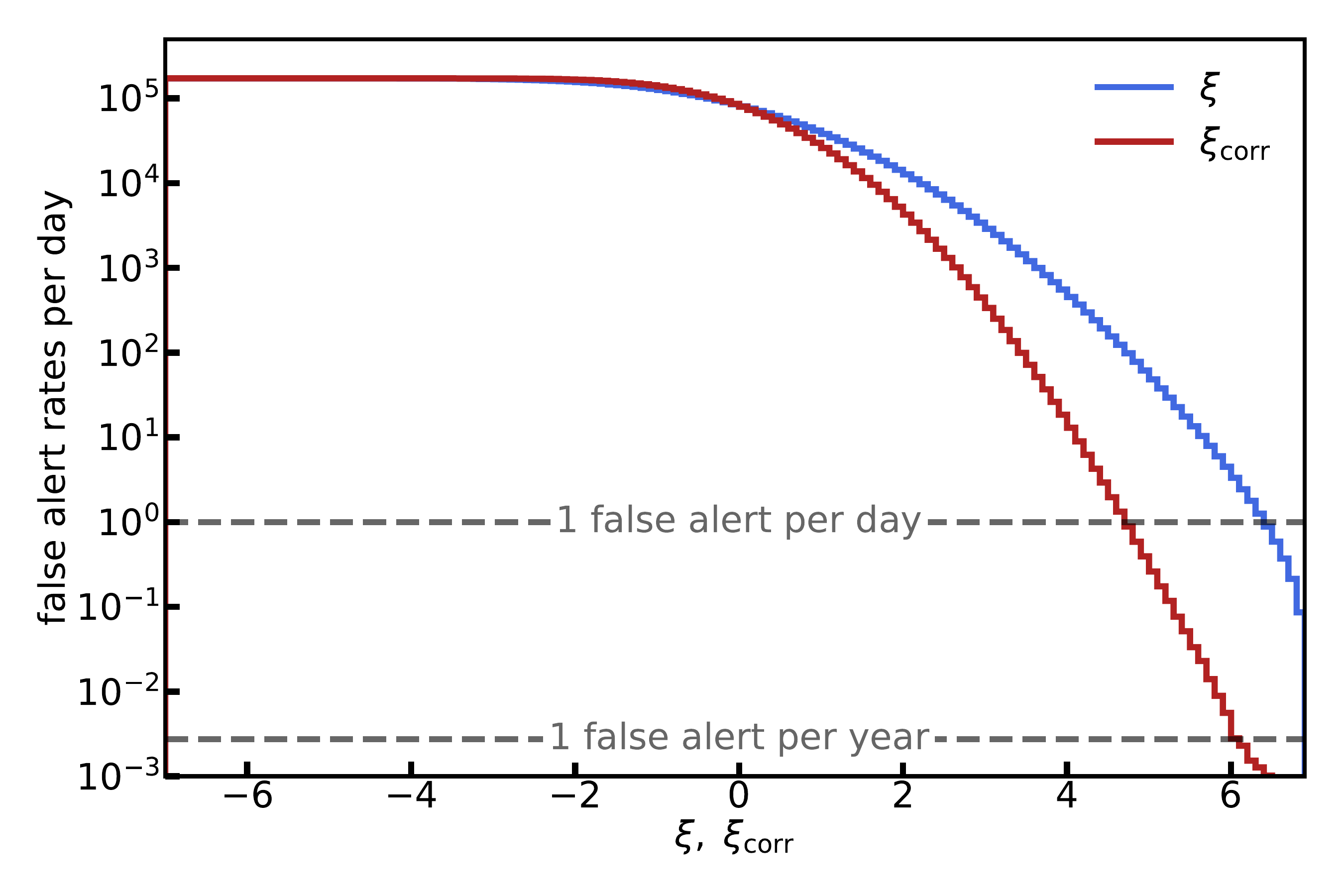}
\caption{\small False alert rate as function of significance proxy threshold. The abscissa shows the cut value on the significance proxy $\xi$ (blue curve) or the muon-corrected significance proxy $\xi_\mathrm{corr}$ (red curve).  The horizontal lines indicate one false alert per day or year, respectively.}
\label{fig:fakealert}
\end{figure}

With the muon-corrected significance proxy $\xi_\mathrm{corr}$ defined, one can determine the false alert rate as function of the significance proxies $\xi$ and $\xi_\mathrm{corr}$. Fig.~\ref{fig:fakealert} shows that the false alert rate can be reduced by a factor of $\approx 400$ for a significance proxy of 6, by applying the muon correction.                        

A slightly less effective atmospheric muon correction is already incorporated in the data acquisition by transmitting the subset of hits associated with IceCube's simple majority triggers~\citep{Kelley:2015ncf} to the supernova data acquisition system. The method has allowed us to lower the alert thresholds and reduce the number of false-positive alerts. For example, the SNEWS alert efficiency for potential supernovae in the Large Magellanic Cloud rose from 12\,\% to 82\,\% while meeting the SNEWS requirement that alerts are sent with a frequency of less than one alert per 14 days.

Lower threshold alerts are issued by the supernova data acquisition system at a rate of about 10 times per day.  
In this sense, the analysis discussed in this paper is not strictly "blind". However, the recalculation of the atmospheric muon-corrected data offline uses a much wider range of triggers and hits. We opted for an unblinding procedure to minimize the influence of prior knowledge by  restricting the initial studies to a data set with  $\xi_\mathrm{corr} < 7$.

\section{Simulation}\label{s:simulation}
A GEANT4-based simulation of the interaction of individual supernova neutrinos in the ice and a computationally optimized tracking~\citep{Schwanekamp:2022ybd} of individual Cherenkov photons that can be run on graphical processing units was used to determine IceCube's effective volume for supernova detection. Calibration measurements with light sources in the ice ~\citep{IceCube:2016zyt,tc-14-2537-2020} and a dust logger~\citep{journal-of-glaciology-2013} allow one to fit the depth-, position- and direction-dependent photon absorption and scattering lengths of the ice. The uncertainties in these measurements lead to a range of possible ice models. The model used in this paper incorporates position dependent scattering and absorption coefficients as well as an observed anisotropic attenuation effect aligned with the local flow of the ice~\citep{Rongen_2021,tc-2022-174}.

Other important uncertainties arise from  photon tracking in the presence of Mie scattering, optical module sensitivities, as well as from neutrino cross section uncertainties, though these are sizable only for interactions on $^{16}$O and $^{18}$O~\citep{IceCube:2011cwc}. The effective volume per optical module was determined by injecting $1.4 \times 10^9$ 
positrons of 10\,MeV energy with random directions and random positions inside a sphere with radius 250\,m around every optical module along every string. 

\begin{figure}[ht]
\centering
\includegraphics[angle=0,width=0.48\textwidth]{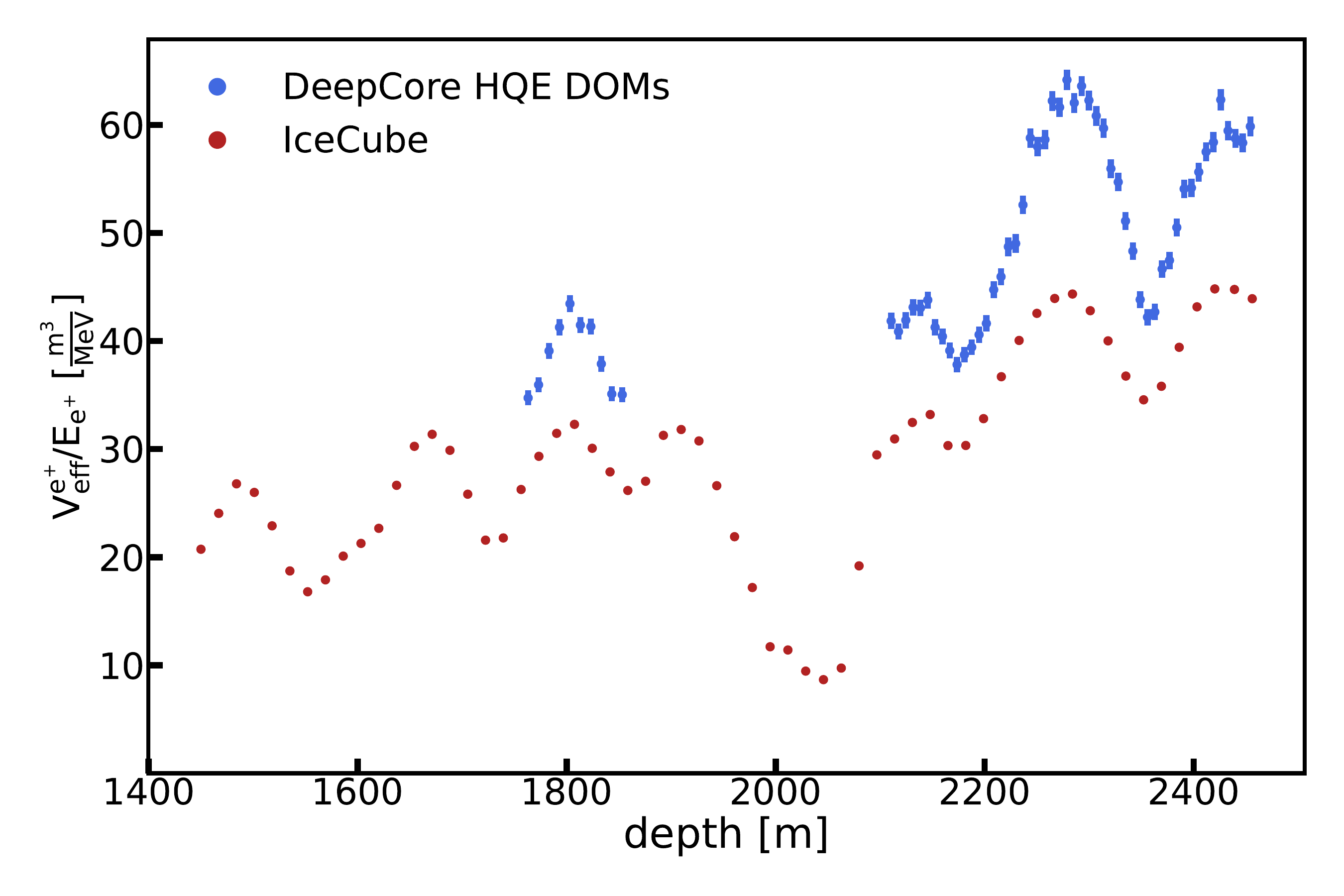}
\caption{$V^\mathrm{eff}_\mathrm{e^+}/E_\mathrm{e^+}$ as function of depth. The variations mirror the depth-dependent absorption coefficient. Note the main dust layer between 1950 and 2100\,m, which corresponds to a glacial maximum 60-70 thousand years ago.
The results are given for standard efficiency DOMs (red) and high efficiency DeepCore DOMs (blue).}\label{fig:Veff}
\end{figure}

Fig.~\ref{fig:Veff} shows the energy-independent quantity $V^\mathrm{eff}_\mathrm{e^+}/E_\mathrm{e^+}$, which was determined from the fraction of
positrons that generated photoelectrons at the cathode surface as function of depth. The $\approx 35\,\%$ higher quantum efficiency of the photomultipliers in the high density DeepCore sub-detector, installed in two ice regions below and above the main dust layer, is apparent. The effective volume scales linearly with the optical module sensitivities. 

\begin{figure}
\includegraphics[angle=0,width=0.48\textwidth]{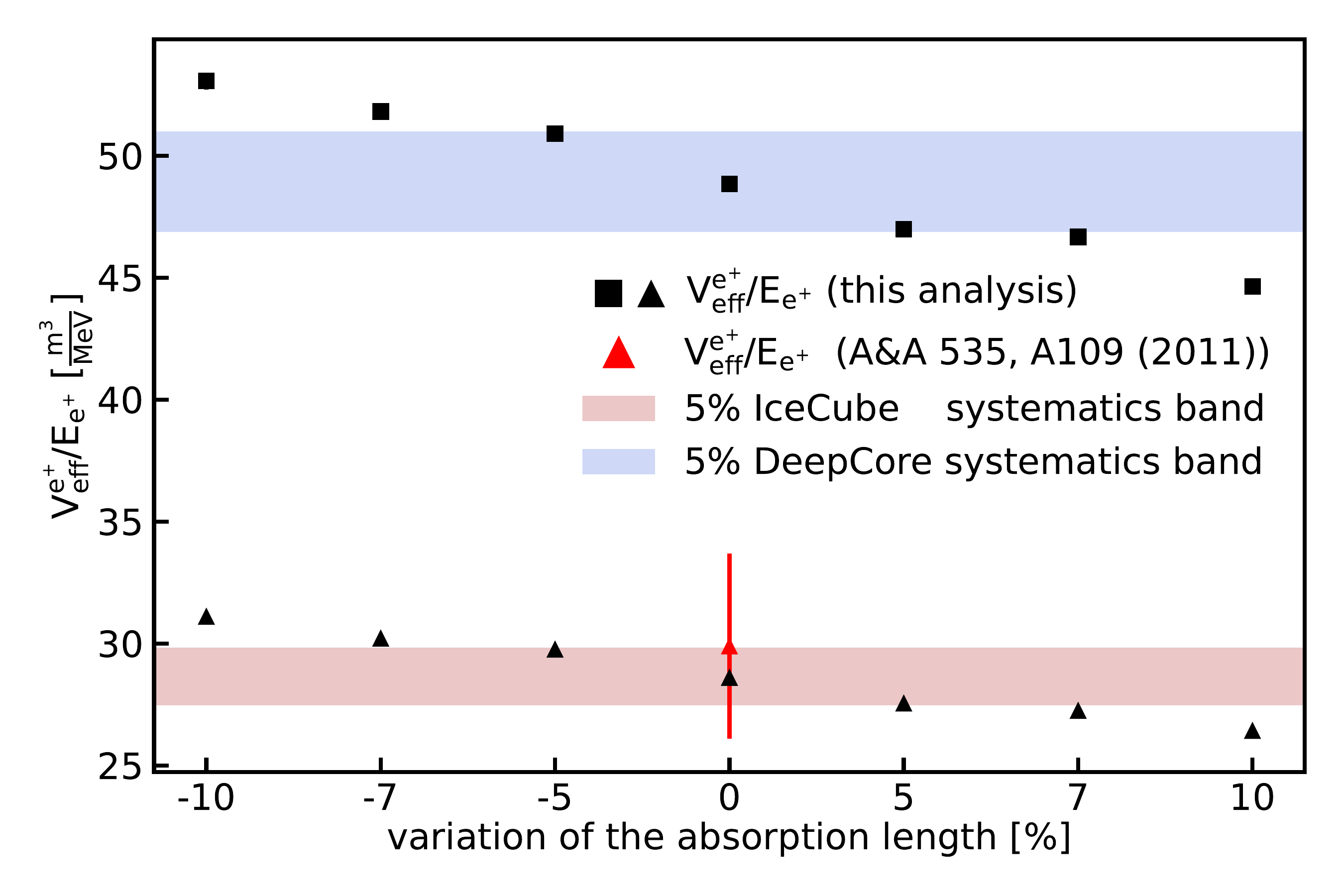}
\caption{\small Left: Systematic uncertainties due to absorption in the ice for IceCube (red) and DeepCore (blue). 
The dependence on the scattering is very small (not shown, see text). 
For comparison, the published value~\citep{IceCube:2011cwc} (in red) is shown including its uncertainty. The upper and lower bands correspond to  5\,\% uncertainties  in the absorption lengths.}\label{fig:Veffsyst}
\end{figure}

While a lot of effort has gone into in-situ calibrations of the ice properties~\citep{IceCube:2013llx,tc-2022-174},  uncertainties remain. The ice density is known to better than 0.2\,\%. The uncertainties on the scattering length, $\lambda_\mathrm{s}$, and absorption length, $\lambda_\mathrm{a}$, are presently estimated at 5\,\% each~\citep{icecubecollaboration2023measurement}. Fig.~\ref{fig:Veffsyst}  shows the result of studies with Monte Carlo samples of $10^7$ generated positrons each, where $\lambda_\mathrm{a}$ was varied within $\pm 10$\,\%. 
A quantitative evaluation shows a strong correlation between the effective volume uncertainty and the change in absorption length with 
$\Delta(V^\mathrm{eff}_\mathrm{e^+}/E_\mathrm{e^+})= -0.7\; (-0.81)\frac{\mathrm{m^2}}{\mathrm{MeV}}\cdot \Delta \lambda_\mathrm{a} \pm 0.02\; (0.04)\frac{\mathrm{m^3}}{\mathrm{MeV}}$ for IceCube (DeepCore). The correlation with the scattering length, on the other hand, is very small: $\Delta(V^\mathrm{eff}_\mathrm{e^+}/E_\mathrm{e^+})=0.037\; (-0.018)\frac{\mathrm{m^2}}{\mathrm{MeV}}\cdot \Delta \lambda_\mathrm{s} \pm 0.015\;(0.037)\frac{\mathrm{m^3}}{\mathrm{MeV}}$ for IceCube (DeepCore). The color bands in Fig.~\ref{fig:Veffsyst} reflect the 5\,\% absorption coefficient uncertainty for IceCube and DeepCore. The result published in ~\citep{IceCube:2011cwc} is  compatible with the new determination based on a much improved understanding of the ice properties. 

Ice properties are not the only source of detector-related uncertainties (see Table~\ref{tab:systematics}). For example, the absolute DOM efficiency in-situ is presently known to 10\,\%. 
In addition, there are uncertainties on the cross sections. Neutrino interactions with oxygen are poorly known; however, they only play a role at neutrino energies beyond 20\,MeV (see Tab.~\ref{tab:systematics}).
Their contributions for 8.8\,$\mathrm{M}_\odot$ progenitor~\citep{Hudepohl:2009tyy} and black hole forming~\citep{Sumiyoshi:2007pp} models are estimated to be 1\,\% and 14\,\%, respectively.

We also studied potential uncertainties due to neutrino oscillations in the Earth matter. These become relevant when comparing the results of detectors at locations with different neutrino path lengths in the Earth or when the supernova position is unknown. The effect on the observed IceCube rate was studied as function of energy and incoming direction. The range of uncertainty for the low 
8.8\,$\mathrm{M}_\odot$ model is given in Table~\ref{tab:systematics}. The uncertainty decreases once the position of the supernova is known. 
\begin{table}[h]
    \centering
    \begin{tabular}{c|c}
    \hline
    \hline
Source of  & 	      estimated relative \\
systematic uncertainty & uncertainty [\%]\\
\hline
\hline
rate deviation in sliding average window & $\pm$ 1.6 \\	
ice density as function of depth &	 $\pm$ 0.2\\
mean $\mathrm{e}^\pm$ track length in ice &  $\pm$ 5 \\
ice properties & [-3.6, 4.1]\\
DOM efficiency & $\pm 10$\\
%effective volume 	& [-13.4, 14.6]\\
artificial dead time  &  $\pm$ 3 \\
cross sections ($\mathrm{e}^+\mathrm{p}$, $\mathrm{e}^-\mathrm{p}$, $\mathrm{e\,O}$ scattering) & < $\pm$ 1, < $\pm$ 1,  $\pm$ 0.2 \\
angle-dependent Earth oscillation & [-0.2,4.9]\\
\hline
\end{tabular}
\caption{Summary of systematic uncertainties on $V_\mathrm{e^+}^\mathrm{eff}$ for the Hüdepohl model~\cite{Hudepohl:2009tyy}. The uncertainties on the oxygen cross sections and angle dependent Earth oscillations are substantially higher in models with larger and more variable neutrino energies. Combining the systematic ice and DOM efficiencies linearly and then in quadrature with the other uncertainties, one obtains an upper error of 16.2\,\% and a lower error of -15.0\,\%.}
    \label{tab:systematics}
\end{table}

Since simulating events with the GEANT4-based Monte Carlo is computationally expensive, we calibrated a parameterized simulation program with these results. The fast Monte Carlo provides access to a large number of supernova  models using time-dependent tables of luminosities, average neutrino energies and spectral shape parameters. The simulation also incorporates various oscillation mechanisms and is capable of injecting signal events into the data stream. One can do without a detailed time-dependent noise and atmospheric muon simulation by using random data sampled over the data taking period.

\begin{figure}[ht]
\centering
\includegraphics[angle=0,width=0.49\textwidth]{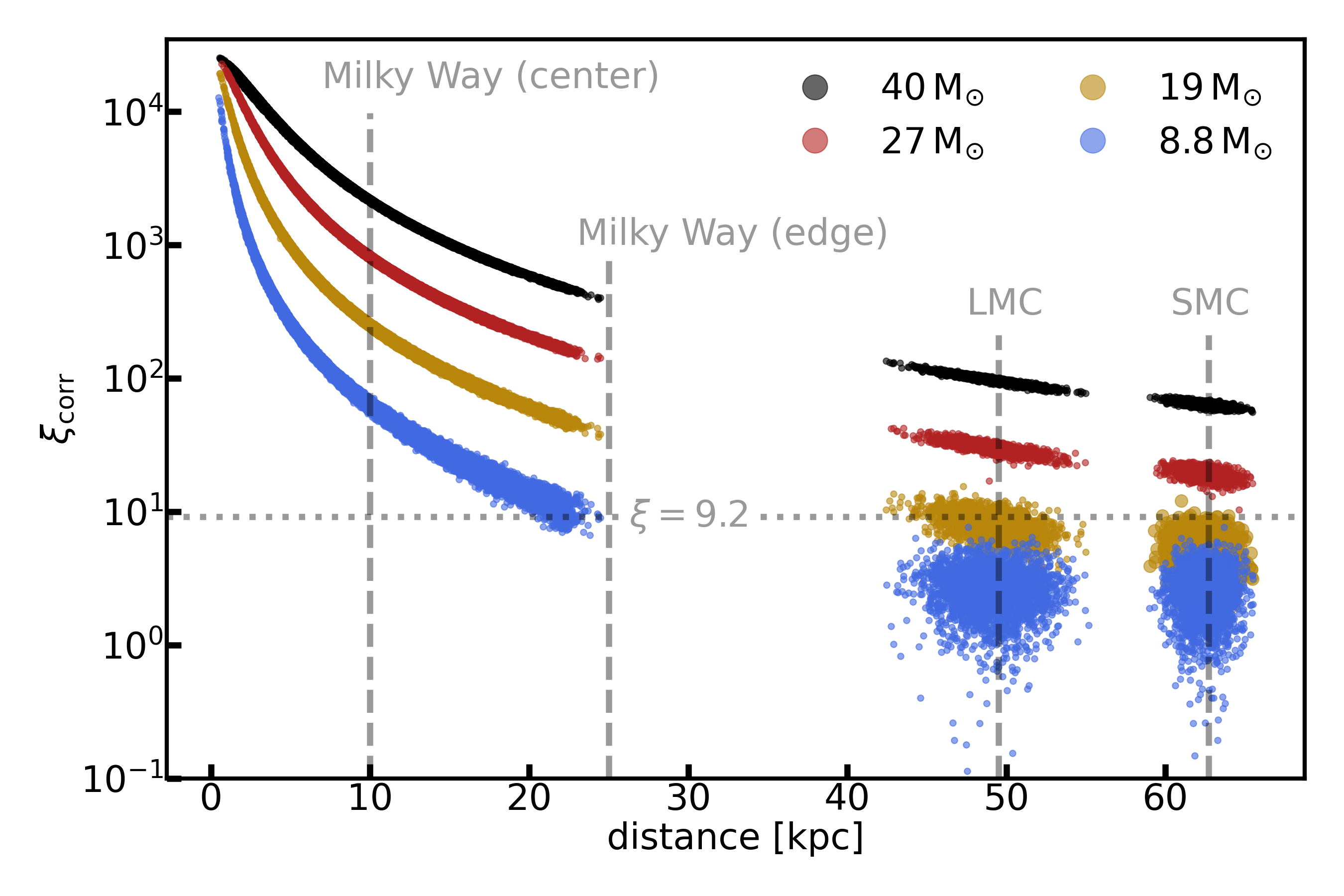}\quad\quad
\caption{Significance proxy versus distance for four models in 0.5\,s bins. The analysis threshold is indicated by a horizontal line.   }\label{fig:signidistance}
\end{figure}

\begin{figure}[ht]
\centering
\includegraphics[width=0.49\textwidth]{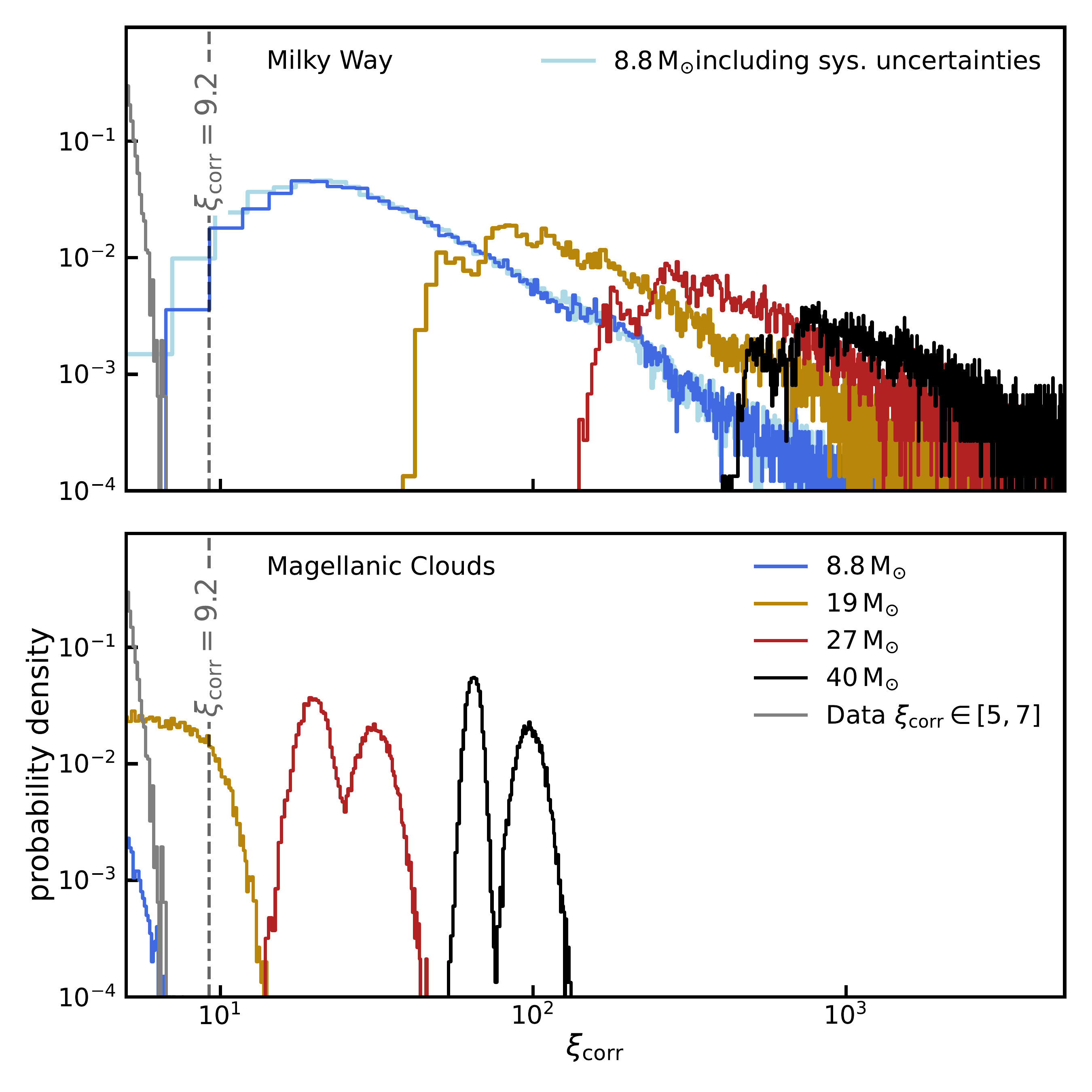}
\caption{Probability densities in our Galaxy (upper plot) and in the Magellanic Clouds (lower plot) for the four models considered. The observed significance proxy $\xi_\mathrm{corr}\in[5,7]$ is also shown. The effect of adding the systematic detector uncertainty is shown for the 8.8\,M$_\odot$ progenitor model in the upper plot (light blue). The overlapping distributions for the Small and Large Magellanic Clouds can not be distinguished for the 8.8\,$\mathrm{M}_\odot$ model. }\label{fig:cumulated}
\end{figure}

Fig.~\ref{fig:signidistance} shows the significance proxy versus the distance for the four selected models. The CCSN distances were simulated to follow the progenitor distribution of ~\cite{Ahlers:2009ae}. 
Fig.~\ref{fig:cumulated} shows the  probability densities as function of $\xi_\mathrm{corr}$ in our Galaxy within 25\,kpc and the Magellanic Clouds for the four models. The data are shown in the range $\xi_\mathrm{corr} \in[5-7]$. The effect of varying the assumed progenitor distance distribution is small (not shown). 

Large uncertainties in the modeling of supernovae may remain even if an optical counterpart can be studied in detail. This is also true for the complex neutrino oscillation effects in the core of the developing supernova.  Addressing these uncertainties goes beyond the scope of this paper.
%
%\vspace{1.0cm}
\section{Results on the Galactic CCSN search} \label{s:Iresults}
\begin{figure}[ht]
\centering
\includegraphics[angle=0,width=0.47\textwidth]{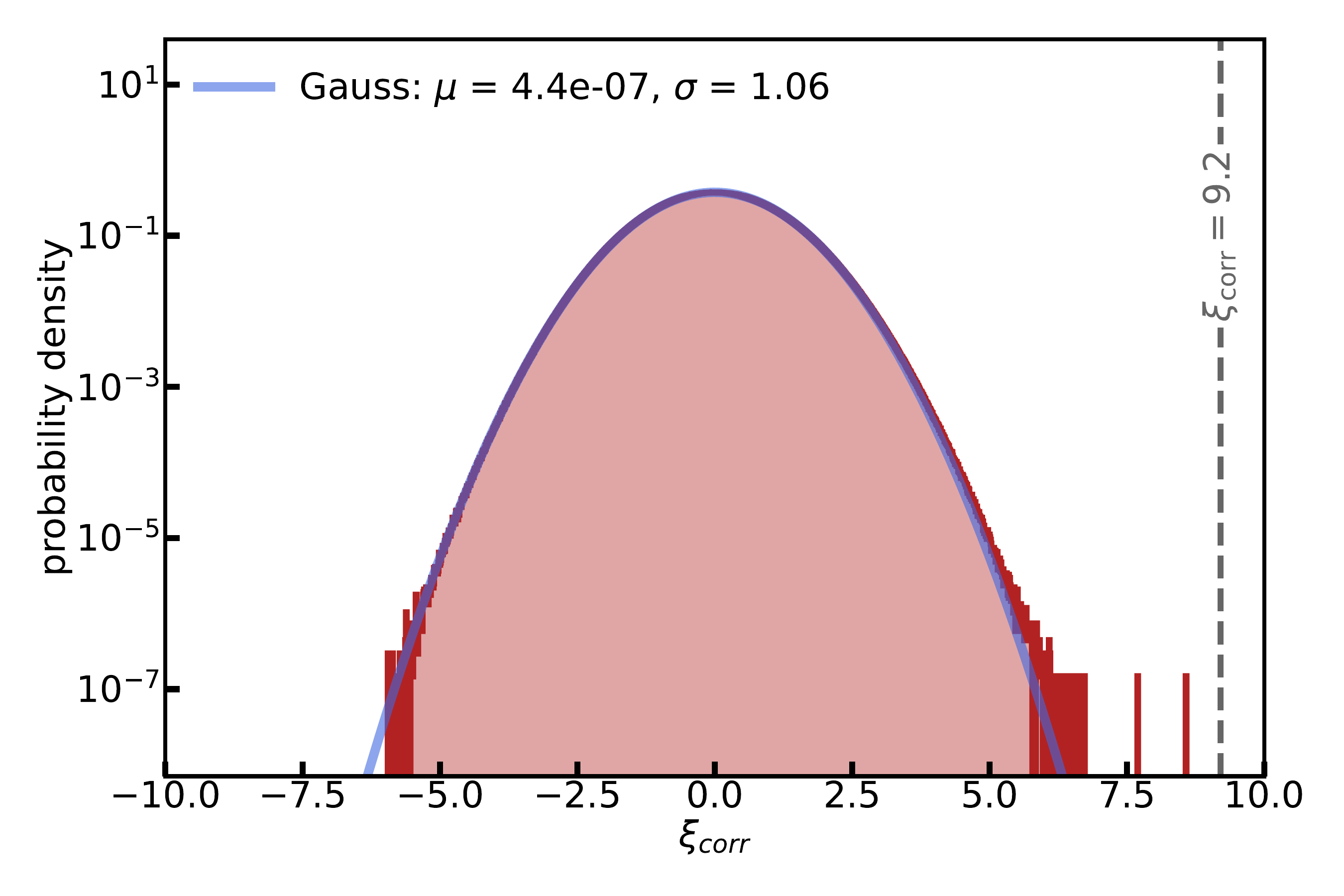}
\caption{Measured $\xi_\mathrm{corr}$ distributions. The fit to a normal distribution (blue line) indicates a deviation from the expectation of the central limit theorem at large positive values.}
\label{fig:closeup}
\end{figure}

From the probability density distributions in Fig.~\ref{fig:cumulated}, including systematic uncertainties, 
we determined the potential signal region by requiring that 99\,\% of all CCSNe in our galaxy for the lightest progenitor studied in this analysis should be detected. For the complete IceCube detector, including the systematic effects in Table~\ref{tab:systematics}, the signal region is defined by $\xi_\mathrm{corr} > 9.2$.
The data with $\xi_\mathrm{corr}> 7$ were then unblinded and the subthreshold range  $\xi_\mathrm{corr}\in [7,9.2]$ and the signal range $\xi_\mathrm{corr} > 9.2$ were investigated. We present in Fig.~\ref{fig:closeup} the accumulated result using the the $\xi_\mathrm{corr}$ distribution. No event in the signal region $\xi_\mathrm{corr}> 9.2$ is found. However, two events entered the subthreshold range, with $\xi_\mathrm{corr} = 7.69$ and  $\xi_\mathrm{corr} = 8.57$. Both events are close in time to failed runs. The estimated background assuming a power law distribution~\citep{1964Pareto} in the blinded region $\xi_\mathrm{corr}> 7$ amounts to $1.2 \pm 0.8$ events.  

A 90\,\% C.L.upper limit was determined using the Feldman-Cousins method~\citep{Feldman_1998}, conservatively assuming no background and the equivalent of 10.735 years of continuous data taking. The limit  
\begin{eqnarray}
N_{90\,\%}/yr &=& 0.23 / yr 
\end{eqnarray}
covers supernovae within 25\,kpc distance, including those that are optically hidden or failed to explode.

As can be seen from Fig.~\ref{fig:cumulated}, only high neutrino luminosity supernovae can be detected in the Small and Large Magellanic Clouds by IceCube alone. However, the situation improves if the burst time is known from external sources, such as other neutrino detectors, a gravitational wave detector, or an astronomical observation with electromagnetic waves.  Unfortunately, the progenitor mass does not uniquely define the observable neutrino flux. We therefore choose to define a progenitor mass-independent measure that is proportional to the observed rate of hits in IceCube. It scales roughly with the third power of the neutrino energy $E_{\nu}$ and depends on the spectral shape that is defined in this analysis by the numerical parameter $\alpha$~\citep{Keil:2002in}. We introduce the quantity
\begin{eqnarray}
X &:=& \int dt \mathcal{L}_\mathrm{SN}^\nu(t)\times \, \langle E_{\nu}(t)\rangle^2\,\times\,\frac{(2+\alpha(t))\cdot (3+\alpha(t))} {(1+\alpha(t)^2)}
\end{eqnarray}
 to set a lower limit for the observation of supernovae in the Magellanic Clouds that fulfill the condition $ X \ge 2.67 \times 10^{61}$\,MeV$^3$. Among the models investigated in this paper, the requirement is satisfied for the 27\,$\mathrm{M}_\odot$ model~\citep{Burrows:2020qrp} and the black hole model~\citep{Sumiyoshi:2007pp}. 

The search discussed in this paper was optimized for the random occurrence of a supernova in the Milky Way and its dwarf galaxy companions. In principle, the sensitivity would be higher for phenomena that occur at fixed frequencies, such as a ``neutrino pulsar''~\citep{Mushtukov:2017ubg}.  Checking the data quality is another reason to study the data set in the frequency domain.
Lomb-Scargle periodograms
%~\citep{bib:lomb,
 %bib:scargle} 
 were used to investigate the data sample in frequency space, up to Nyquist frequencies of 1 Hz~\citep{Fritz:2021btf,Fritz:2022yzy}. With the exception of a signature of the diurnal seasonal oscillation of the muon rate and artifacts from the run transitions and alias effects at high frequency, no significant signal was found. 

\section{Conclusions}\label{s:conclusion}
A search for neutrinos from core-collapse supernovae in the Milky Way and dwarf galaxy companions using IceCube
data taken between April 17, 2008 and December 31, 2019 was performed. The period covers the equivalent of 10.735 years of uninterrupted data taking. With the cuts defined in this analysis and for distances smaller than 25\,kpc, IceCube has the sensitivity to detect 99\,\% of all Galactic core-collapse supernovae with neutrino fluxes equal to or higher than that of the conservative $8.8\,\mathrm{M}_\odot$ progenitor model. No candidate event was found and a 90\,\% C.L. upper limit on the rate of core-collapse 
supernovae out to distances of $\approx$ 25\,kpc was determined to be 0.23/yr. This limit can be extended to the Magellanic Clouds for models that fulfill the condition $ X \ge 2.67 \times 10^{61} $\,MeV$^3$, with the progenitor mass independent measure $X$ defined in section~\ref{s:Iresults}.

As part of the approved
IceCube Upgrade, multi-PMT modules~\citep{IceCube:2021eij} will be deployed and low-noise wavelength-shifting
sensors~\citep{Bastian-Querner:2021uqv} will be tested. They  have the potential to increase the distance reach~\citep{LozanoMariscal:2021the} and will substantially improve the
spectral sensitivity.
\vspace{2cm}
\section*{Acknowledgements}
The IceCube collaboration acknowledges the significant contribution to this manuscript from Alexander Fritz. The authors gratefully acknowledge the support from the following gencies: 

USA {\textendash} U.S. National Science Foundation-Office of Polar Programs,
U.S. National Science Foundation-Physics Division,
U.S. National Science Foundation-EPSCoR,
Wisconsin Alumni Research Foundation,
Center for High Throughput Computing (CHTC) at the University of Wisconsin{\textendash}Madison,
Open Science Grid (OSG),
Advanced Cyberinfrastructure Coordination Ecosystem: Services {\&} Support (ACCESS),
Frontera computing project at the Texas Advanced Computing Center,
U.S. Department of Energy-National Energy Research Scientific Computing Center,
Particle astrophysics research computing center at the University of Maryland,
Institute for Cyber-Enabled Research at Michigan State University,
and Astroparticle physics computational facility at Marquette University;
Belgium {\textendash} Funds for Scientific Research (FRS-FNRS and FWO),
FWO Odysseus and Big Science programmes,
and Belgian Federal Science Policy Office (Belspo);
Germany {\textendash} Bundesministerium f{\"u}r Bildung und Forschung (BMBF),
Deutsche Forschungsgemeinschaft (DFG),
Helmholtz Alliance for Astroparticle Physics (HAP),
Initiative and Networking Fund of the Helmholtz Association,
Deutsches Elektronen Synchrotron (DESY),
and High Performance Computing cluster of the RWTH Aachen;
Sweden {\textendash} Swedish Research Council,
Swedish Polar Research Secretariat,
Swedish National Infrastructure for Computing (SNIC),
and Knut and Alice Wallenberg Foundation;
European Union {\textendash} EGI Advanced Computing for research;
Australia {\textendash} Australian Research Council;
Canada {\textendash} Natural Sciences and Engineering Research Council of Canada,
Calcul Qu{\'e}bec, Compute Ontario, Canada Foundation for Innovation, WestGrid, and Compute Canada;
Denmark {\textendash} Villum Fonden, Carlsberg Foundation, and European Commission;
New Zealand {\textendash} Marsden Fund;
Japan {\textendash} Japan Society for Promotion of Science (JSPS)
and Institute for Global Prominent Research (IGPR) of Chiba University;
Korea {\textendash} National Research Foundation of Korea (NRF);
Switzerland {\textendash} Swiss National Science Foundation (SNSF);
United Kingdom {\textendash} Department of Physics, University of Oxford.

\newpage
\bibliography{references}{}

\begin{thebibliography}{}
\expandafter\ifx\csname natexlab\endcsname\relax\def\natexlab#1{#1}\fi
\providecommand{\url}[1]{\href{#1}{#1}}
\providecommand{\dodoi}[1]{doi:~\href{http://doi.org/#1}{\nolinkurl{#1}}}
\providecommand{\doeprint}[1]{\href{http://ascl.net/#1}{\nolinkurl{http://ascl.net/#1}}}
\providecommand{\doarXiv}[1]{\href{https://arxiv.org/abs/#1}{\nolinkurl{https://arxiv.org/abs/#1}}}

\bibitem[{Aartsen {et~al.}(2013{\natexlab{a}})}]{journal-of-glaciology-2013}
Aartsen, M.~G., {et~al.} 2013{\natexlab{a}}, Journal of Glaciology, 59,
  1117–1128, \dodoi{10.3189/2013JoG13J068}

\bibitem[{Aartsen {et~al.}(2013{\natexlab{b}})}]{IceCube:2013llx}
---. 2013{\natexlab{b}}, Nucl. Instrum. Meth. A, 711, 73,
  \dodoi{10.1016/j.nima.2013.01.054}

\bibitem[{Aartsen {et~al.}(2017)}]{IceCube:2016zyt}
---. 2017, JINST, 12, P03012, \dodoi{10.1088/1748-0221/12/03/P03012}

\bibitem[{Aartsen {et~al.}(2020{\natexlab{a}})}]{Aartsen_2020}
---. 2020{\natexlab{a}}, JINST, 15, P06032,
  \dodoi{10.1088/1748-0221/15/06/P06032}

\bibitem[{Aartsen {et~al.}(2020{\natexlab{b}})}]{IceCube:2019acm}
---. 2020{\natexlab{b}}, Astrophys. J., 890, 111,
  \dodoi{10.3847/1538-4357/ab564b}

\bibitem[{Abbasi {et~al.}(2010)}]{IceCube:2010dpc}
Abbasi, R., {et~al.} 2010, Nucl. Instrum. Meth. A, 618, 139,
  \dodoi{10.1016/j.nima.2010.03.102}

\bibitem[{Abbasi {et~al.}(2011)}]{IceCube:2011cwc}
---. 2011, Astron. Astrophys., 535, A109, \dodoi{10.1051/0004-6361/201117810e}

\bibitem[{Abbasi {et~al.}(2012)}]{IceCube:2011ucd}
---. 2012, Astropart. Phys., 35, 615,
  \dodoi{10.1016/j.astropartphys.2012.01.004}

\bibitem[{Abbasi {et~al.}(2021)}]{IceCube:2021eij}
---. 2021, PoS, ICRC2021, 1070, \dodoi{10.22323/1.395.1070}

\bibitem[{Abbasi {et~al.}(2022)}]{tc-2022-174}
---. 2022, The Cryosphere Discussions, 2022, 1, \dodoi{10.5194/tc-2022-174}

\bibitem[{Abbasi {et~al.}(2023)}]{icecubecollaboration2023measurement}
---. 2023, Measurement of Atmospheric Neutrino Mixing with Improved IceCube
  DeepCore Calibration and Data Processing.
\newblock \doarXiv{2304.12236}

\bibitem[{Abbott {et~al.}(2016)}]{LIGOScientific:2016aoc}
Abbott, B.~P., {et~al.} 2016, Phys. Rev. Lett., 116, 061102,
  \dodoi{10.1103/PhysRevLett.116.061102}

\bibitem[{Abe {et~al.}(2016)}]{Super-Kamiokande:2016kji}
Abe, K., {et~al.} 2016, Astropart. Phys., 81, 39,
  \dodoi{10.1016/j.astropartphys.2016.04.003}

\bibitem[{Abe {et~al.}(2022)}]{KamLAND:2022sqb}
Abe, S., {et~al.} 2022, Astrophys. J., 934, 85,
  \dodoi{10.3847/1538-4357/ac7a3f}

\bibitem[{Abi {et~al.}(2021)}]{Abi_2021}
Abi, B., {et~al.} 2021, Eur. Phys. J. C, 81,
  \dodoi{10.1140/epjc/s10052-021-09166-w}

\bibitem[{Acero {et~al.}(2021)}]{NOvA:2021zhv}
Acero, M.~A., {et~al.} 2021, Phys. Rev. D, 104, 063024,
  \dodoi{10.1103/PhysRevD.104.063024}

\bibitem[{Adams {et~al.}(2013)Adams, Kochanek, Beacom, Vagins, \&
  Stanek}]{Adams:2013ana}
Adams, S.~M., Kochanek, C.~S., Beacom, J.~F., Vagins, M.~R., \& Stanek, K.~Z.
  2013, Astrophys. J., 778, 164, \dodoi{10.1088/0004-637X/778/2/164}

\bibitem[{Agafonova {et~al.}(2015)}]{LVD:2014uzr}
Agafonova, N.~Y., {et~al.} 2015, Astrophys. J., 802, 47,
  \dodoi{10.1088/0004-637X/802/1/47}

\bibitem[{Aharmim {et~al.}(2011)}]{SNO:2010noh}
Aharmim, B., {et~al.} 2011, Astrophys. J., 728, 83,
  \dodoi{10.1088/0004-637X/728/2/83}

\bibitem[{Ahlers {et~al.}(2009)Ahlers, Mertsch, \& Sarkar}]{Ahlers:2009ae}
Ahlers, M., Mertsch, P., \& Sarkar, S. 2009, Phys. Rev. D, 80, 123017,
  \dodoi{10.1103/PhysRevD.80.123017}

\bibitem[{Ahrens {et~al.}(2002)}]{AMANDA:2001htp}
Ahrens, J., {et~al.} 2002, Astropart. Phys., 16, 345,
  \dodoi{10.1016/S0927-6505(01)00154-2}

\bibitem[{Aiello {et~al.}(2022)}]{KM3NeT:2021oaa}
Aiello, S., {et~al.} 2022, Eur. Phys. J. C, 82, 317,
  \dodoi{10.1140/epjc/s10052-022-10137-y}

\bibitem[{Al~Kharusi {et~al.}(2021)}]{SNEWS:2020tbu}
Al~Kharusi, S., {et~al.} 2021, New J. Phys., 23, 031201,
  \dodoi{10.1088/1367-2630/abde33}

\bibitem[{Alekseev {et~al.}(1987)Alekseev, Alekseeva, Volchenko, \&
  Krivosheina}]{Alekseev:1987ej}
Alekseev, E.~N., Alekseeva, L.~N., Volchenko, V.~I., \& Krivosheina, I.~V.
  1987, JETP Lett., 45, 589

\bibitem[{Alsabti \& Murdin(2017)}]{2017hsn..book.....A}
Alsabti, A.~W., \& Murdin, P., eds. 2017, {Handbook of Supernovae} (Springer
  International Publishing), \dodoi{10.1007/978-3-319-20794-0}

\bibitem[{Ambrosio {et~al.}(2004)}]{MACRO:2004fzv}
Ambrosio, M., {et~al.} 2004, Eur. Phys. J. C, 37, 265,
  \dodoi{10.1140/epjc/s2004-01981-3}

\bibitem[{Antonioli {et~al.}(2004)}]{Antonioli:2004zb}
Antonioli, P., {et~al.} 2004, New J. Phys., 6, 114,
  \dodoi{10.1088/1367-2630/6/1/114}

\bibitem[{Bastian-Querner {et~al.}(2022)}]{Bastian-Querner:2021uqv}
Bastian-Querner, B., {et~al.} 2022, Sensors, 22, 1385,
  \dodoi{10.3390/s22041385}

\bibitem[{Baxter {et~al.}(2022)}]{baxter2021snewpy}
Baxter, A.~L., {et~al.} 2022, Astrophys. J, 925, 107,
  \dodoi{10.3847/1538-4357/ac350f}

\bibitem[{Bionta {et~al.}(1987)}]{Bionta:1987qt}
Bionta, R.~M., {et~al.} 1987, Phys. Rev. Lett., 58, 1494,
  \dodoi{10.1103/PhysRevLett.58.1494}

\bibitem[{Bollig {et~al.}(2021)}]{Bollig:2020phc}
Bollig, R., {et~al.} 2021, Astrophys. J., 915, 28,
  \dodoi{10.3847/1538-4357/abf82e}

\bibitem[{Brdar {et~al.}(2018)Brdar, Lindner, \& Xu}]{Brdar_2018}
Brdar, V., Lindner, M., \& Xu, X.-J. 2018, JCAP, 2018, 025,
  \dodoi{10.1088/1475-7516/2018/04/025}

\bibitem[{Burrows \& Vartanyan(2021)}]{Burrows:2020qrp}
Burrows, A., \& Vartanyan, D. 2021, Nature, 589, 29,
  \dodoi{10.1038/s41586-020-03059-w}

\bibitem[{{Coleiro, A.} {et~al.}(2020){Coleiro, A.}, {Molla, M. Colomer},
  {Dornic, D.}, {Lincetto, M.}, \& {Kulikovskiy, V.}}]{refId0}
{Coleiro, A.}, {Molla, M. Colomer}, {Dornic, D.}, {Lincetto, M.}, \&
  {Kulikovskiy, V.} 2020, Eur. Phys. J. C, 80, 856,
  \dodoi{10.1140/epjc/s10052-020-8407-7}

\bibitem[{D\'\i{}az-Rodr\'\i{}guez {et~al.}(2021)D\'\i{}az-Rodr\'\i{}guez,
  Murphy, Williams, Dalcanton, \& Dolphin}]{Diaz-Rodriguez:2021xja}
D\'\i{}az-Rodr\'\i{}guez, M., Murphy, J.~W., Williams, B.~F., Dalcanton, J.~J.,
  \& Dolphin, A.~E. 2021, Monthly Notices of the Royal Astronomical Society,
  506, 781, \dodoi{10.1093/mnras/stab1800}

\bibitem[{Feldman \& Cousins(1998)}]{Feldman_1998}
Feldman, G.~J., \& Cousins, R.~D. 1998, Phys. Rev. D, 57, 3873,
  \dodoi{10.1103/physrevd.57.3873}

\bibitem[{Fritz(2022)}]{Fritz:2022yzy}
Fritz, A. 2022, PhD thesis, Mainz U., \dodoi{10.25358/openscience-7825}

\bibitem[{Fritz \& Kappesser(2021)}]{Fritz:2021btf}
Fritz, A., \& Kappesser, D. 2021.
\newblock \doarXiv{2108.05138}

\bibitem[{Gullin {et~al.}(2022)Gullin, O\textquoteright{}Connor, Wang, \&
  Tseng}]{Gullin:2021hfv}
Gullin, S., O\textquoteright{}Connor, E.~P., Wang, J.-S., \& Tseng, J. 2022,
  Astrophys. J., 926, 212, \dodoi{10.3847/1538-4357/ac4420}

\bibitem[{Hakobyan {et~al.}(2015)Hakobyan, Karapetyan, Barkhudaryan, Mamon,
  Kunth, Petrosian, Adibekyan, Aramyan, \& Turatto}]{10.1093/mnras/stv2853}
Hakobyan, A.~A., Karapetyan, A.~G., Barkhudaryan, L.~V., {et~al.} 2015, Monthly
  Notices of the Royal Astronomical Society, 456, 2848,
  \dodoi{10.1093/mnras/stv2853}

\bibitem[{Halzen {et~al.}(1996)Halzen, Jacobsen, \& Zas}]{Halzen:1995ex}
Halzen, F., Jacobsen, J.~E., \& Zas, E. 1996, Phys. Rev. D, 53, 7359,
  \dodoi{10.1103/PhysRevD.53.7359}

\bibitem[{Heereman~von Zuydtwyck(2015)}]{HeeremanvonZuydtwyck:2015mbs}
Heereman~von Zuydtwyck, D.~F. 2015, PhD thesis, U. Brussels (main).
\newblock
  \url{https://difusion.ulb.ac.be/vufind/Record/ULB-DIPOT:oai:dipot.ulb.ac.be:2013/209179/Holdings}

\bibitem[{Hirata {et~al.}(1987)}]{Kamiokande-II:1987idp}
Hirata, K., {et~al.} 1987, Phys. Rev. Lett., 58, 1490,
  \dodoi{10.1103/PhysRevLett.58.1490}

\bibitem[{Hirata {et~al.}(1988)}]{Hirata:1988ad}
Hirata, K.~S., {et~al.} 1988, Phys. Rev. D, 38, 448,
  \dodoi{10.1103/PhysRevD.38.448}

\bibitem[{H{\"u}depohl {et~al.}(2010)H{\"u}depohl, M{\"u}ller, Janka, Marek, \&
  Raffelt}]{Hudepohl:2009tyy}
H{\"u}depohl, L., M{\"u}ller, B., Janka, H.~T., Marek, A., \& Raffelt, G.~G.
  2010, Phys. Rev. Lett., 104, 251101, \dodoi{10.1103/PhysRevLett.104.251101}

\bibitem[{Ikeda {et~al.}(2007)}]{Super-Kamiokande:2007zsl}
Ikeda, M., {et~al.} 2007, Astrophys. J., 669, 519, \dodoi{10.1086/521547}

\bibitem[{Janka(2012)}]{JankaReview}
Janka, H.-T. 2012, Annu. Rev. Nucl. Part. Sci., 62, 407,
  \dodoi{10.1146/annurev-nucl-102711-094901}

\bibitem[{Keil {et~al.}(2003)Keil, Raffelt, \& Janka}]{Keil:2002in}
Keil, M.~T., Raffelt, G.~G., \& Janka, H.-T. 2003, Astrophys. J., 590, 971,
  \dodoi{10.1086/375130}

\bibitem[{Kelley(2015)}]{Kelley:2015ncf}
Kelley, J.~L. 2015, AIP Conf. Proc., 1630, 154, \dodoi{10.1063/1.4902795}

\bibitem[{Kersten \& Smirnov(2016)}]{Kersten:2015kio}
Kersten, J., \& Smirnov, A.~Y. 2016, Eur. Phys. J. C, 76, 339,
  \dodoi{10.1140/epjc/s10052-016-4187-5}

\bibitem[{Kochanek {et~al.}(2008)Kochanek, Beacom, Kistler, Prieto, Stanek,
  Thompson, \& Yuksel}]{Kochanek:2008mp}
Kochanek, C.~S., Beacom, J.~F., Kistler, M.~D., {et~al.} 2008, Astrophys. J.,
  684, 1336, \dodoi{10.1086/590053}

\bibitem[{Kotake {et~al.}(2006)Kotake, Sato, \& Takahashi}]{Kotake:2005zn}
Kotake, K., Sato, K., \& Takahashi, K. 2006, Rept. Prog. Phys., 69, 971,
  \dodoi{10.1088/0034-4885/69/4/R03}

\bibitem[{Lang {et~al.}(2016)Lang, McCabe, Reichard, Selvi, \&
  Tamborra}]{Lang_2016}
Lang, R.~F., McCabe, C., Reichard, S., Selvi, M., \& Tamborra, I. 2016, Phys.
  Rev. D, 94, \dodoi{10.1103/physrevd.94.103009}

\bibitem[{Lattimer \& Prakash(2001)}]{Lattimer:2000nx}
Lattimer, J.~M., \& Prakash, M. 2001, Astrophys. J., 550, 426,
  \dodoi{10.1086/319702}

\bibitem[{Linzer \& Scholberg(2019)}]{PhysRevD.100.103005}
Linzer, N.~B., \& Scholberg, K. 2019, Phys. Rev. D, 100, 103005,
  \dodoi{10.1103/PhysRevD.100.103005}

\bibitem[{Lozano~Mariscal {et~al.}(2021)}]{LozanoMariscal:2021the}
Lozano~Mariscal, C.~J., {et~al.} 2021, Eur. Phys. J. C, 81, 1058,
  \dodoi{10.1140/epjc/s10052-021-09809-y}

\bibitem[{Lund \& Kneller(2013)}]{PhysRevD.88.023008}
Lund, T., \& Kneller, J.~P. 2013, Phys. Rev. D, 88, 023008,
  \dodoi{10.1103/PhysRevD.88.023008}

\bibitem[{Maggi {et~al.}(2016)}]{Maggi:2015uga}
Maggi, P., {et~al.} 2016, Astron. Astrophys., 585, A162,
  \dodoi{10.1051/0004-6361/201526932}

\bibitem[{Malmenbeck \& O'Sullivan(2019)}]{Malmenbeck:2019giw}
Malmenbeck, F., \& O'Sullivan, E. 2019, in {36th International Cosmic Ray
  Conference}.
\newblock \doarXiv{1909.00886}

\bibitem[{Meyer(2010)}]{Meyer_2010}
Meyer, H.~O. 2010, Europhysics Letters, 89, 58001,
  \dodoi{10.1209/0295-5075/89/58001}

\bibitem[{Migenda {et~al.}(2021)Migenda, Cartwright, Kneale, Malek,
  Schnellbach, \& Stone}]{Migenda:2021hnl}
Migenda, J., Cartwright, S., Kneale, L., {et~al.} 2021, J. Open Source Softw.,
  6, 2877, \dodoi{10.21105/joss.02877}

\bibitem[{Mirizzi {et~al.}(2016)}]{rivistareview}
Mirizzi, A., {et~al.} 2016, La Rivista del Nuovo Cimento, 39, 1–112,
  \dodoi{10.1393/ncr/i2016-10120-8}

\bibitem[{{Moore} \& {Kasliwal}(2019)}]{2019NatAs...3..109M}
{Moore}, A.~M., \& {Kasliwal}, M.~M. 2019, Nature Astronomy, 3, 109,
  \dodoi{10.1038/s41550-018-0675-x}

\bibitem[{Mushtukov {et~al.}(2018)}]{Mushtukov:2017ubg}
Mushtukov, A.~A., {et~al.} 2018, Monthly Notices of the Royal Astronomical
  Society, 474, 5425, \dodoi{10.1093/mnras/stx2905}

\bibitem[{Neustadt {et~al.}(2021)}]{Neustadt:2021jjt}
Neustadt, J.~M., {et~al.} 2021, Monthly Notices of the Royal Astronomical
  Society, 508, 516, \dodoi{10.1093/mnras/stab2605}

\bibitem[{Novoseltsev {et~al.}(2020)}]{Novoseltsev:2019gdt}
Novoseltsev, Y.~F., {et~al.} 2020, Astropart. Phys., 117, 102404,
  \dodoi{10.1016/j.astropartphys.2019.102404}

\bibitem[{Novoseltsev {et~al.}(2022)}]{Novoseltsev:2022lmd}
---. 2022, J. Exp. Theor. Phys., 134, 390, \dodoi{10.1134/S1063776122040082}

\bibitem[{O'Connor(2017)}]{O’Connor2017}
O'Connor, E. 2017, The Core-Collapse Supernova-Black Hole Connection, ed. A.~W.
  Alsabti \& P.~Murdin (Springer International Publishing),
  \dodoi{10.1007/978-3-319-20794-0_129-1}

\bibitem[{O'Connor {et~al.}(2018)}]{OConnor:2018sti}
O'Connor, E., {et~al.} 2018, J. Phys. G, 45, 104001,
  \dodoi{10.1088/1361-6471/aadeae}

\bibitem[{{Pareto}(1964)}]{1964Pareto}
{Pareto}, V. 1964, {Cours d'Économie Politique: Nouvelle édition par G.-H.
  Bousquet et G. Busino}, 299–345

\bibitem[{Pryor {et~al.}(1988)Pryor, Roos, \& Webster}]{Pryor:1987tz}
Pryor, C., Roos, C.~E., \& Webster, M.~S. 1988, Astrophys. J., 329, 335,
  \dodoi{10.1086/166381}

\bibitem[{Rongen {et~al.}(2020)Rongen, Bay, \& Blot}]{tc-14-2537-2020}
Rongen, M., Bay, R.~C., \& Blot, S. 2020, The Cryosphere, 14, 2537,
  \dodoi{10.5194/tc-14-2537-2020}

\bibitem[{Rongen \& Chirkin(2021)}]{Rongen_2021}
Rongen, M., \& Chirkin, D. 2021, Journal of Instrumentation, 16, C09014,
  \dodoi{10.1088/1748-0221/16/09/C09014}

\bibitem[{Rosso(2021)}]{Gallo_Rosso_2021}
Rosso, A.~G. 2021, JCAP, 2021, 046, \dodoi{10.1088/1475-7516/2021/06/046}

\bibitem[{Rozwadowska {et~al.}(2021)Rozwadowska, Vissani, \&
  Cappellaro}]{Rozwadowska:2020nab}
Rozwadowska, K., Vissani, F., \& Cappellaro, E. 2021, New Astron., 83, 101498,
  \dodoi{10.1016/j.newast.2020.101498}

\bibitem[{Rumleskie \& Virtue(2020)}]{Rumleskie:2020iip}
Rumleskie, J., \& Virtue, C. 2020, J. Phys. Conf. Ser., 1342, 012135,
  \dodoi{10.1088/1742-6596/1342/1/012135}

\bibitem[{Scholberg {et~al.}(2022)}]{snowglobes}
Scholberg, K., {et~al.} 2022, SNOwGLoBES,
  \url{https://github.com/SNOwGLoBES/snowglobes},  GitHub

\bibitem[{Schwanekamp {et~al.}(2022)}]{Schwanekamp:2022ybd}
Schwanekamp, H., {et~al.} 2022, Comput. Softw. Big Sci., 6, 4,
  \dodoi{10.1007/s41781-022-00080-8}

\bibitem[{Shen {et~al.}(1998)Shen, Toki, Oyamatsu, \& Sumiyoshi}]{SHEN1998435}
Shen, H., Toki, H., Oyamatsu, K., \& Sumiyoshi, K. 1998, Nuclear Physics A,
  637, 435, \dodoi{https://doi.org/10.1016/S0375-9474(98)00236-X}

\bibitem[{Smartt(2015)}]{Smartt_2015}
Smartt, S.~J. 2015, Publications of the Astronomical Society of Australia, 32,
  \dodoi{10.1017/pasa.2015.17}

\bibitem[{Sumiyoshi {et~al.}(2007)Sumiyoshi, Yamada, \&
  Suzuki}]{Sumiyoshi:2007pp}
Sumiyoshi, K., Yamada, S., \& Suzuki, H. 2007, Astrophys. J., 667, 382,
  \dodoi{10.1086/520876}

\bibitem[{Vigorito {et~al.}(2021)Vigorito, Bruno, Fulgione, \&
  Molinario}]{Vigorito:2021sgy}
Vigorito, C.~F., Bruno, G., Fulgione, W., \& Molinario, A. 2021, PoS, ICRC2021,
  1111, \dodoi{10.22323/1.395.1111}

\bibitem[{{Vink}(2020)}]{2020pesr.book.....V}
{Vink}, J. 2020, {Physics and Evolution of Supernova Remnants},
  \dodoi{10.1007/978-3-030-55231-2}

\end{thebibliography}
\bibliographystyle{aasjournal}
\end{document}